\documentclass[twocolumn]{aastex63}
\usepackage{hyperref}
\usepackage{bm}
\usepackage{multirow}
\usepackage{gensymb}
\usepackage{enumitem}
\usepackage{tabularx}
\usepackage[flushleft]{threeparttable}
\usepackage[]{color}

\usepackage{natbib}
\usepackage{float}
\usepackage{rotating}
\usepackage{amsfonts,amsmath,amssymb}
\usepackage{graphicx}
\usepackage{color}
\usepackage{xcolor}
\usepackage{url}
\usepackage{letltxmacro}

\newcommand{\vth}{v_{\text{th}}}

\newcommand{\xstar}{\textsc{\scriptsize XSTAR}}

\newcommand{\athena}{\textsc{Athena\scriptsize ++ }}
\newcommand{\E}[1]{\left\langle#1\right\rangle}

\def\hep0{{\rm HEP_{0}}}
\def\fmtxi{$M(t,\xi)$~}

\def\<{\,\langle\langle}
\def\>{\,\rangle\rangle}

\def\MSUN{\rm M_{\odot}}

\def\CIVdbl{{\rm C~}\kern 0.1em{\sc iv}~$\lambda\lambda 1548, 1550$} 
\def\OVIIIi{\hbox{{\rm O}\kern 0.1em{\sc viii}}}
\def\SiXIVi{\hbox{{\rm Si}\kern 0.1em{\sc viii}}}
\def\OVIII{\hbox{{\rm O}\kern 0.1em{\sc viii}~{\rm Ly}$\alpha$}}
\def\SiXIV{\hbox{{\rm Si}\kern 0.1em{\sc viii}~{\rm Ly}$\beta$}}
\def\FeXXV{\hbox{{\rm Fe}\kern 0.1em{\sc xxv}}}
\def\FeXXVI{\hbox{{\rm Fe}\kern 0.1em{\sc xxvi}}}
\def\FeXXVK{\hbox{{\rm Fe}\kern 0.1em{\sc xxv}~{\rm K}$\alpha$}}
\def\FeXXVIK{\hbox{{\rm Fe}\kern 0.1em{\sc xxvi}~{\rm K}$\alpha$}}

\newcommand{\beq}{\begin{equation}}
\newcommand{\seq}{\end{equation}}

\renewcommand{\v}[1]{\ensuremath{\mathbf{#1}}} 
\newcommand{\gv}[1]{\ensuremath{\mbox{\boldmath$ #1 $}}} 
\newcommand{\uv}[1]{\ensuremath{\mathbf{\hat{#1}}}} 
\renewcommand{\d}[2]{\frac{d #1}{d #2}} 

\newcommand{\f}{\frac}  
\newcommand{\e}[1]{\exp\left( #1 \right)} 
\def\fourbb{$4\times10^{5}~{\rm K}$ BB}
\def\eightbb{$8\times10^{5}~{\rm K}$ BB}
\def\milbb{$10^{6}~{\rm K}$ BB}
\usepackage{letltxmacro,amsmath}
\LetLtxMacro{\originaleqref}{\eqref}
\renewcommand{\eqref}{Eq.~\originaleqref}

\defcitealias{Dyda17}{D17}
\defcitealias{Dannen19}{D19}
\defcitealias{Dannen20}{D20}
\defcitealias{Waters21}{Waters et al. 2021}
\defcitealias{Begelman83}{BMS83}
\defcitealias{CAK}{CAK}
\defcitealias{SK90}{SK90}

\submitjournal{ApJ}
\shorttitle{Line driving in the cold and warm wind regimes}
\shortauthors{Dannen et al.}
\begin{document}
\title{On the transition from efficient to inefficient line-driving 
in irradiated flows}

\correspondingauthor{Randall Dannen}
\author[0000-0002-5160-8716]{Randall Dannen}
\email{randall.dannen@unlv.edu}
\affiliation{Department of Physics \& Astronomy \\
University of Nevada, Las Vegas \\
4505 S. Maryland Pkwy \\
Las Vegas, NV, 89154-4002, USA}

\author[0000-0002-6336-5125]{Daniel Proga}
\affiliation{Department of Physics \& Astronomy \\
University of Nevada, Las Vegas \\
4505 S. Maryland Pkwy \\
Las Vegas, NV, 89154-4002, USA}

\author[0000-0002-5205-9472]{Tim Waters}
\affiliation{Theoretical Division, Los Alamos National Laboratory}


\begin{abstract}
Observations of ionized AGN outflows have provided compelling evidence 
that the radiation field transfers both momentum and energy to the plasma. 
At parsec scale distances in AGN, energy transfer can dominate, in which 
case the only force needed to launch an outflow is that from gas pressure. 
Much closer to the black hole, gravity dominates thermal energy due 
to insufficient heating by the radiation and the gas is in the so-called 
`cold' regime. Only magnetic or radiation forces can then lead 
to outflow, but it is unclear at what temperature and ionization state 
the radiation force weakens, as these properties depend on the spectral energy 
distribution (SED).  In this work, we survey the parameter space of radiation forces 
due to spectral lines resulting from blackbody SEDs with varying temperatures 
in the range $\sim 10^4 - 10^6$~K to identify the radiation temperature 
at which line-driving begins to lose efficiency.
We find that the temperature $\lesssim4\times10^5$~K  marks 
the transition to inefficient line driving.
We also self-consistently compute the heating and cooling balance to estimate the gas temperature, 
so that our parameter survey covers the transition where thermal driving goes from negligible 
to comparable to line driving.  
We summarize a large set of hydrodynamical simulations of radial flows to illustrate 
how the wind properties change during the transition and
the dependence of these properties on the assumed SED and governing flow parameters. 
\end{abstract}

\section{Introduction} \label{sec:intro}

Radiation can play a significant role in launching and accelerating mass outflows 
in active galactic nuclei (AGN) and in other astrophysical systems such as OB stars and 
cataclysmic variables (CVs). On parsec scale distances in AGN, 
where there is a weak gravitational potential, energy transfer from radiation 
to the plasma can result in considerable heating, allowing 
thermal driving to produce an outflow. On scales where radiation heating 
is insufficient because gravitational potential energy dominates thermal energy, 
it can be a good approximation to initially neglect the gas pressure force altogether; 
this was done in the line-driven wind model of 
\citeauthor{CAK} (\citeyear{CAK}; hereafter \citetalias{CAK}) 
and in the magneto-centrifugal wind model of \citet{BlandfordPayne}, 
two classic examples of analytic `cold' wind solutions.

Determining the relative role of thermal and radiation driving 
is challenging because it requires accurate treatment of a non-trivial coupling between 
electromagnetic radiation and matter (i.e., the gas opacity and emissivity) from 
the underlying spectral energy distribution (SED) of the radiation
\citep[for textbook reviews, see, e.g., ][and references therein]{MihalasMihalas84, Castor07}.
In some applications,  accurate treatment may seem unnecessary because
one could justify ignoring the radiation force
by referring to the rule of thumb that “radiation can heat (cool), but frequently it 
finds it difficult to push” \cite{Shu92}. However, as many rules of thumb, 
it does not apply everywhere and radiation can push gas when 
the total opacity of the gas, $\kappa_{\rm tot}$, 
is dominated by the contribution from photon scattering.
OB stars and CVs are examples of objects with strong winds and negligible thermal driving.
In the upper atmospheres and winds of these objects, 
$\kappa_{\rm tot}$ is dominated  by contributions from spectral line transitions, which mostly scatter photons,  
hence their winds are driven by the line force, $F_{\rm rad, l}$ \citepalias[as in][]{CAK}.  
It is possible that this line driving mechanism is responsible for producing 
supersonic outflows in AGNs as suggested by numerous observations 
\citep[e.g.,][]{Foltz87, Srianand02, Ganguly03, Gupta03, North06, Bowler14, Lu18, MasRibas19}
and some theoretical work \citep[e.g.,][]{Mushotzky72, Arav94, MCGV, PSK00, P07}. 

In OB stars, the so-called overionization problem 
\citep[e.g.,][]{Arav94, MCGV, Krolik99, dKB95, PSK00}
is not an issue because these objects emit radiation similar to blackbodies (BBs) with low temperatures. 
Therefore, ionizing photons are produced in small numbers and with 
low energies compared to CVs or AGN, which have much broader UV continuum 
spectra due to the presence of a multi-temperature BB disk. 
We note that SEDs for CVs have ionizing properties that fall somewhere 
between those of OB stars and AGN, as they do possess a soft X-ray 
contribution from a boundary layer  \citep[e.g.,][]{Frank02}.

The overionization problem becomes a major concern when modeling AGN outflows, however.  
Computing the ionization balance is necessary, as it serves as a crucial observational constraint, 
while also determining the magnitude of the line force.
There are two seemingly contradictory  observational results to consider. 
On the one hand, AGN are characterized by very high fluxes in both the UV and X-rays 
and on the other hand, they possess strong spectral lines from moderately and highly ionized species.
The likely explanation for how the ionization balance remains hospitable to the observed line 
formation is self-shielding, in which the radiation penetrating outflowing gas 
is filtered through denser regions of plasma near the base of the wind
(\citealp[e.g.,][]{MCGV, PSK00, PK04}, but see also \citealp{Sim10, Higginbottom14}).
Developing self-consistent models, therefore, requires coupling hydrodynamical 
(`hydro' for short) simulations of the gas dynamics with ionization balance calculations. 

{To identify when exactly overionization becomes a problem for line-driving, 
it is necessary to survey the parameter space of the line force resulting from BB SEDs spanning a broad range of temperatures spanning $2\times 10^4$~K to $10^6$~K, while also self-consistently computing the heating 
and cooling balance to estimate the gas temperature.  This need makes it 
impractical to employ the most sophisticated approach for coupling matter 
and radiation, namely a multi-group treatment of radiation hydrodynamics (RHD), 
as described by \citet[][see also references therein]{Jiang22}. In the latter,
performing calculations with a large number of energy bins as needed to accurately 
sample SEDs is still not entirely feasible.  
Indeed, most RHD simulations rely on a `gray' treatment of radiation, i.e. using
only the Planck, energy, and flux averaged continuum opacities.

In recent years, we have developed a complementary approach for accounting for both 
line driving and various radiative heating and cooling processes,
including Compton processes, that does not require solving the equations of multi-group RHD
(\citeauthor{Dyda17} \citeyear{Dyda17}, \citeauthor{Dannen19} \citeyear{Dannen19}; 
\citetalias{Dyda17} and \citetalias{Dannen19}, respectively).
In any particular hydro run, we assume a fixed SED, for which we can construct a grid of 
photoionization models covering all possible ionization states and temperatures that could 
be encountered in that run.  We then self-consistently couple these ionization balance calculations 
with the hydrodynamics by interpolating from the grid of photoionization models to evaluate the 
source terms resulting from the net cooling function and the radiation force.
Our approach is computationally inexpensive, which is important for exploring 
a wide range of parameter space.

This methodology has already proven useful for providing new insights 
into the problem of AGN outflows.
For example, in \citeauthor{Dannen20} (\citeyear{Dannen20}; hereafter \citetalias{Dannen20}),
we studied radial winds driven by AGN SEDs, 
finding that when winds originate at large radii, thermal driving dominants over line driving.
Our analysis there showed that at the wind base, despite the low degree of ionization 
and an abundance of spectral lines, the line force is weak. 
This weakness is caused by large optical depths in lines that suppress the line force.
On the other hand, further into the wind, the gas is highly 
ionized and as a result, there are fewer lines available, decreasing the overall line force.
The few lines are a manifestation of the overionization problem.
In addition, the few lines mean that line cooling is inefficient 
and the gas experiences runaway heating to various degrees depending 
on the Compton temperature \citepalias[e.g.,][]{Dyda17, Waters21}.  
Thus, overionization not only suppresses line driving but also enhances thermal driving.

While overionization is a well-known concern, the issue of excessively 
high optical depth in lines at the wind base is new. 
In this paper, we thoroughly study this previously unexplored regime of parameter space.
The objective is to identify regions where the radiation force dominates over thermal driving 
in determining the mass loss rate or terminal velocity (or both).
In \S\,\ref{sec:method}, we describe our methods to arrive at wind solutions 
by solving the hydrodynamic equations including heating and cooling and line driving. 
In \S\,\ref{sec:results}, we present the results of our numerical calculations. 
We conclude and discuss our results in \S\,\ref{sec:conclusions} and note that our paper 
also includes two appendices: Appendix \ref{appendix-teq-v-fm} contains 
a summary of the photoionization results that we utilize in our wind models, while
Appendix \ref{appendix-relations} contains scaling relationships between parameters 
that are helpful to determine different physical regimes for line driving.
\par


\section{Methods} \label{sec:method}
As mentioned above, we described our framework for modeling irradiated outflows 
from a variety of objects (e.g., stars, accretion disks around stars as well as black holes) 
in a series of papers \citepalias{Dyda17, Dannen19, Dannen20}.
This framework was built upon several previous studies of gas outflows in
CVs and AGN \citep[e.g.,][]{Proga98,PSK00, P07}.
\par

\subsection{Photoionization calculations}
In \citetalias{Dyda17}, we demonstrated our method for modeling outflows 
resulting from the irradiation of optically thin gas by a radiation field 
with an {\it arbitrary} strength and SED. We used  the photoionization code 
\xstar\footnote{\url{https://heasarc.nasa.gov/lheasoft/xstar/xstar.html}}~\citep{KB01} 
to calculate the radiative heating and cooling rates
($\mathcal{H}$ and $\mathcal{C}$, respectively) as  functions of gas temperature, $T$, 
and gas ionization parameter
\begin{equation}
\xi = (4\pi)^2 \frac{J_X}{n_{\rm H}},
\label{eq:xi-def}
\end{equation}
where $J_X$ is the integrated  mean intensity from 0.1~Ry--1000~Ry 
and $n_{\rm H}$ the hydrogen nucleon number density.
\footnote{When $J_X$ is due to a point source so that the ionizing flux 
is $F_X = 4\pi J_X$, this ionization parameter is the familiar $\xi = 4\pi F_X/n_{\rm H}$.}
We explored several SEDs: 
those due to unobscured and obscured AGN \citep[hereafter AGN1 and AGN2,][]{Mehdipour15}, 
as well as SEDs for hard and soft state X-ray binaries 
(hereafter XRB1 and XRB2, \citealt{Trigo13b}, see Fig.~1 in \citetalias{Dyda17}), 
bremsstrahlung, and blackbody (BB).  We applied our method to study the hydrodynamics 
of 1-D spherical winds heated  by a uniform radiation field using 
the magnetohydrodynamic code \athena \citep{Stone20}. \par

In  our followup work, \citetalias{Dannen19},
we presented the next step in our development 
of a self-consistent method to model astrophysical winds.  
As in \citetalias{Dyda17}, we employed the photoionization code \xstar, 
this time to compute not only $\mathcal{H}(\xi, T)$ and $\mathcal{C}(\xi, T)$ 
but also the radiation force due to spectral lines using the most complete 
and up-to-date line list. For a radial flow, the line force 
${\bf F}_{\rm rad,l} = M(\xi,t){\bf F}_{\rm rad,e}$, 
where ${\bf F}_{\rm rad,e}$ 
is the radiation force due to electron scattering and
$M$ is the famous force multiplier introduced by
\citetalias{CAK}. 
For a given SED, the force multiplier is 
a function of $\xi$ and $T$ and, in addition, 
of an optical depth parameter, $t=\sigma_e \rho \lambda_{\rm Sob}$, where $\sigma_e$
is the mass-scattering coefficient for free electrons, $\rho$ is the mass density 
and $\lambda_{\rm Sob}$ the Sobolev length. 
This length is a measure of how much the gas has accelerated relative to thermal 
velocity and equals to $v_{th}/({d v_l}/d l)$, 
where $v_{th}$ is the thermal velocity while ${d v_l}/d l$ 
is the sight-line velocity gradient.
However, to reduce the number of parameters,
we assumed the temperature dependence can be captured by using
the thermal equilibrium value of $T$ for a given $\xi$ so
that ${M}=M(t,\xi,T_{\rm eq}(\xi))={M}(t,\xi)$
where $T_{\rm eq}(\xi)$ satisfies the equilibrium equation: 
$\mathcal{L(\xi,T_{\rm eq})}\equiv\mathcal{C}(\xi,T_{\rm eq})-\mathcal{H}(\xi,T_{\rm eq})=0$,
$\mathcal{L}$ is a net cooling function.

In this paper, our main focus is to present our hydrodynamical simulations.
However, we also wish to compile the outcomes of our past and present photoionization 
computations into a single location.
Therefore, we showcase a selection of our results
from the photoionization calculations for different SEDs 
in Appendix~\ref{appendix-teq-v-fm} and present a portion of these results  
in Fig.~\ref{fig:fm}.

\begin{figure*}[htb!]
  \centering
  \includegraphics[width=\textwidth]{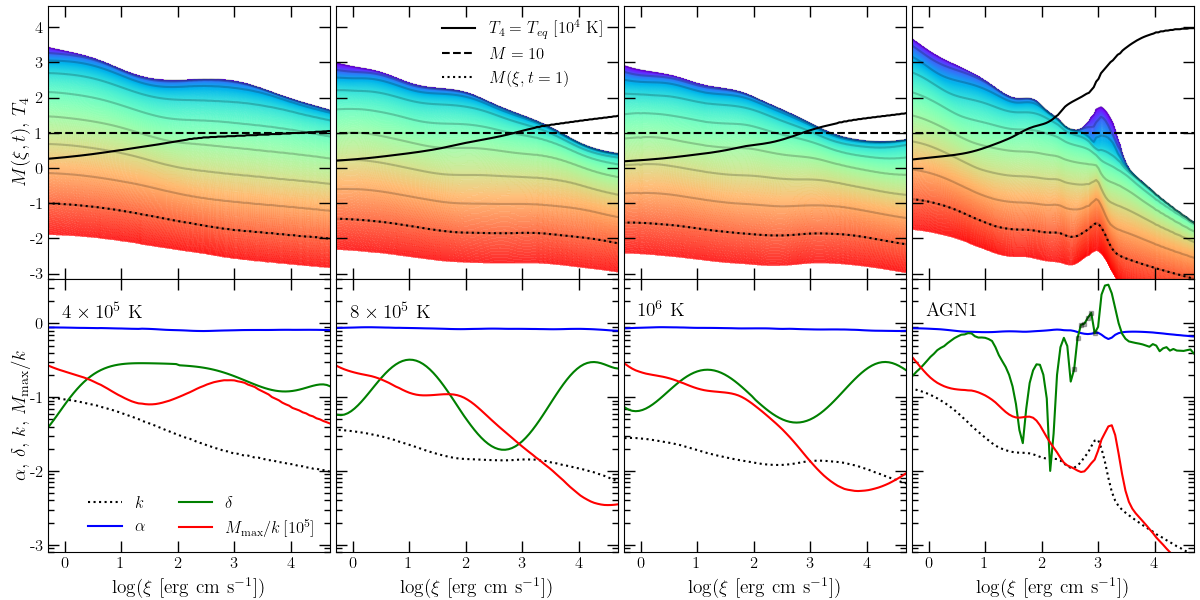}
  \caption{
  Example of results from our photoionization calculations for four SEDs: from left to right, 
  blackbody SEDs with temperatures $4\times 10^{5}$~K, $8\times 10^{5}$~K, $10^{6}$~K, 
  and an unobscured AGN SED \citep[AGN1;][]{Mehdipour15}.
  {\textit{Top panels}}: colormaps of the force multiplier for $t$ decreasing from $\log t = 1$ (red) 
  to $\log t = -8$ (blue). Black dotted lines mark \fmtxi, while gray lines 
  indicate \fmtxi at $\log t= -1, -2, ..., -8$, in ascending order, from bottom to top.
  The horizontal dashed line corresponds to the value \fmtxi$=10$, indicating 
  the threshold at which line driving becomes the primary driving force for outflows.  
  The solid black curve (S-curve) corresponds to the equilibrium temperature, $T_{\rm eq}$, in units of $10^4$~K.
  {\textit{Bottom panels}}: profiles of four parameters characterizing 
  the behavior of the force multiplier: $\alpha(\xi)$, $k(\xi)$, $|\delta(\xi)|$, and $M_{max}(\xi)/k(\xi)$ 
  (blue, black dashed, green, and red curves, respectively). To indicate any negative values of $\delta$, 
  we added green squares to the right panel's green line. We calculate the first three parameters 
  along the $M(t=1, \xi)$ curves to establish a connection between our numerical findings for the force multiplier 
  and the modified version of the formula proposed by \citetalias{CAK}. Finally, we determine
  the fourth parameter, $M_{max}(\xi)/k(\xi)$, using the $M(t=10^{-8}, \xi)$ curve  
  (the highest curve among the curves in the top panels) as a proxy of the maximum force
  multiplier, $M_{max}(\xi)$. See Appendix~\ref{appendix-teq-v-fm} for more details.
   }
  \label{fig:fm}
\end{figure*}

\cite{SK90} presented one of the earliest quantifications of the overionization problem 
using detailed photoionization calculations with \xstar. Assuming a 10~keV bremsstrahlung SED, 
\cite{SK90} found that the force multiplier is a strong function of the photoionization 
parameter for high values of the parameter. In particular, for a given $t$, \fmtxi 
quickly approaches zero for  $\xi>100$. This $\xi$-dependence is an illustration 
of the overionization problem in line-driven winds. The results from \citet{SK90} 
have been applied to model outflows in AGN for over two decades 
\citep[e.g.,][]{PSK00, PK04, Kurosawa09, Nomura13, WP16, Nomura20, Quera-Bofarull21, Wang22}.
However,  the AGN SEDs differ from a bremsstrahlung SED. Therefore, for given $\xi$ and $T$, 
the gas exposed to an AGN radiation field can have different spectral lines
than that exposed to bremsstrahlung radiation.

In \citetalias{Dannen19}, we confirmed that overall the line force is still 
a strong function of $\xi$ for high values of $\xi$ when using AGN SEDs. 
We also found that for a fixed value of $t$, the force multiplier is not a monotonic 
function of $\xi$ (see the right panel in Fig.~\ref{fig:fm} and the first two panels in Fig.~\ref{fig:fm_a}).
While \fmtxi first decreases with $\xi$ for $\log\xi \lesssim 2$ as shown by \cite{SK90}. 
For both AGN1 and AGN2 SEDs, this decrease is not as strong as compared to 
a 10~keV bremsstrahlung SED used by \cite{SK90}.
In addition, we found that for a fixed $t$, \fmtxi can increase again at $\log\xi \approx 1.5$ 
and more prominently at $\log\xi \approx 3$ for AGN1.
The main consequence of this behavior is that the multiplier remains larger
than 1 even at $\log\xi \approx 3$, as shown in the right panel in Fig.~\ref{fig:fm}.
Moreover, we noted that the range $10^2 \lesssim \xi \lesssim 10^3$ is also where gas 
is thermally unstable by the isobaric criterion.
The regime of thermal instability \cite[TI;][]{Field65} corresponds to 
the $\log T$--$\log \xi$ curve having a slope larger than 1 
(see the black solid curve in the top panels in Figs.~\ref{fig:fm}~and~\ref{fig:fm_a}). 

\subsection{Hydrodynamical calculations}

We solve the equations of hydrodynamics with additional source terms to include 
radiative and cooling processes as well as radiation force: 
\begin{equation} \label{eq:hydro-matter}
\frac{\partial \rho}{\partial t} + \nabla\cdot (\rho {\bf v}) =0,
\end{equation} 
\quad
\begin{equation} \label{eq:hydro-mom}
\frac{\partial \rho {\bf v}}{\partial t} 
+ \nabla\cdot (\rho {\bf v}{\bf v} + {\bf P}) = 
-\rho \nabla \Phi + {\bf F}_{\text{rad}},
\end{equation} 
and
\begin{equation} \label{eq:hydro-en}
\frac{\partial E}{\partial t} + \nabla \cdot [(E+p){\bf v}] =
-\rho{\bf v} \cdot \nabla \Phi - \rho \mathcal{L} + 
 {\bf v}\cdot {\bf F}_{\text{rad}},
\end{equation} 
here, $\rho$, $p$, ${\bf v}$, and $\mathcal{E}$) are
the gas mass density, pressure, velocity, and specific internal energy, respectively. 
In addition,
${\bf P} = p\, {\bf I}$ where ${\bf I}$ is the unit tensor, 
$\Phi=-GM/r$ is the gravitational potential due to a body with 
with mass $M$, $E = \rho \mathcal{E} + 1/2 \rho |{\bf v}|^2$ 
is the total energy,
${\bf F}^{\rm rad}=F^{\rm rad}\, \hat{\bm r}$ is the radiation force, 
and $\mathcal{L}$ is the net cooling rate.  We adopt an adiabatic equation of state 
$p=(\gamma-1)\rho \mathcal{E}$ and assume $\gamma = 5/3$ in the cases 
where radiative heating and cooling are present. However, we also consider nearly 
isothermal cases where we set $\gamma = 1.0001$ and $\mathcal{L} = 0$. 
Our time-dependent hydrodynamical simulations are performed in spherical coordinates under spherical symmetry (i.e. assuming radial streamlines). 

We use pre-calculated tables of $\mathcal{L} = \mathcal{L}(\xi,T)$, which is coupled to the hydro using the backward Euler scheme described in \citetalias{Dyda17}.  As opposed to the linear interpolation method employed by \citetalias{Dyda17}, $\mathcal{L}(\xi,T)$ is evaluated using a cubic interpolation method from the GSL library as described by \citet{Waters21}, who found that this increases the accuracy by factor of about five.
In addition, we apply this interpolation scheme 
when evaluating the force multiplier tables, which are arrived at by following the methods presented in \citetalias{Dannen19}.
These grids of photoionization calculations are illustrated in Figure~\ref{fig:fm}.

We initialize the density assuming a hydrostatic atmosphere such 
that $\rho(r) = \rho_{0}\exp{(-{\rm HEP}_{0} ( r/r_{0} - 1 ) )}$ 
with constant temperature set to be  $T_{\rm eq}(\xi_0)$
(for the definition of $\hep0$, see Eq.~\ref{eq:hep} in Appendix~\ref{appendix-relations}). 
The initial velocity for the models that include heating and cooling is set to 0. 
However, for the isothermal models, we apply the velocity field 
$\gv{v}(r) = v_{\rm esc}\sqrt{1 - r_{0}/r}~\uv{r}$.
At the beginning of our simulations, we apply outflow boundary conditions 
at the inner and outer radii and allow the hydro to evolve for ~10-30\% 
of the total run time.  We then pause and restart
the simulation with constant gradient boundary conditions applied at the outer radius.
We also fix the density in the first active zone to the appropriate value as shown 
in Table~\ref{tab:input-sum}. Our standard computational domain 
is defined to occupy the radial range $r_0\leq r \leq 100 r_0$, where  $r_0$ is computed 
based on our assumed value of  $\xi_0$. In Table~\ref{tab:input-sum}, we list
the main  input parameters of our calculations as well as a short summary of how 
we compute $r_0$ and $\rho_0$. Finally, we adopt logarithmic grid spacing such that 
$dr_{i+1}=1.0099dr_{i}$ and use $N_{r}=1020$ grid cells in the radial direction.  

For comparison purposes, we present models that include line-driving implemented using the prescription from \citetalias{CAK} or modified CAK (mCAK hereafter).
These have the force multiplier given by the formula
\begin{equation}\label{eq:mcak}
    M(t,\xi) = k(\xi) t^{-\alpha},
\end{equation}
where $k(\xi)= k_0 \left( \xi_{0} / \xi \right)^{\delta}$
(see Appendix~\ref{appendix-teq-v-fm} for our explanation of this $k-\xi$ scaling).
In these models, whether or not $\delta$ is set to zero, we assume $k_0 = 0.0076$ and $\alpha=0.742$.

\par

\begin{table}[]
\centering
\resizebox{\columnwidth}{!}{
\begin{tabular}{rcccc}
\multicolumn{1}{r|}{$T_{\rm BB}$~[K]} & $T_{0}$~[$10^4$~K] & $\rho_{0}$ [$10^{-11}$~g~cm$^{-3}$] & $r_0$ [$10^{13}$~cm] & $\Xi_{0}$ \\ \hline
\multicolumn{1}{r|}{$4\times10^{5}$}  & 5.51               & 11.1                                & 1.70                 & 28    \\
\multicolumn{1}{r|}{$8\times10^{5}$}  & 4.76               & 8.28                                & 1.95                 & 32    \\
\multicolumn{1}{r|}{$10^{6}$}         & 4.26               & 0.66                                & 1.76                 & 26    \\ \hline
\end{tabular}%
}
\caption{Example of input parameters for  three SED cases 
with ${\rm HEP}_0=5$: $4\times10^5$, $8\times10^5$, and $10^6$~BB SED,
the second, third, and fourth columns, respectively.
For all models, we assumed $M = 10~\MSUN$, $\Gamma = 0.2$, and $\xi_0=80$.
In addition, we set $T_0$ to $T_{eq}$ corresponding to $\xi_0$ for a given SED. 
Therefore, the inner radius and the density at the base depend only on $\hep0$ 
and we compute them using the following expressions 
$r_0 = \left(1 - \Gamma \right) GM~{\text{HEP}_0}^{-1} c_{s,0}^{-2}$ 
(see Eq.~\ref{eq:hep}) and
$\rho_0 = \mu m_{p} L_{\rm Edd}\Gamma\xi_{0}^{-1} r_{0}^{-2}$. For reference, 
we also list $\Xi_0$ (see Eq.~\ref{eq:Xi-def}).
We explored models with $\hep0$ ranging from 5 to 500 
(see Table~\ref{tab:result-sum}).
}
\label{tab:input-sum}
\end{table}

\begin{table*}[]
\centering
\begin{tabular}{rcccc}
Model                                   & HEP & $\dot{M}$ [$M_\odot$~yr$^{-1}$] &  $\eta_{\rm wind}$  & comment \\ \hline
\multirow{7}{*}{$4\times10^{5}$~K}      & 5   &  $2.99\times10^{-7}$   &  $3.55\times10^{-2}$  & steady state \\
                                        & 10  &  $9.78\times10^{-8}$   &  $1.47\times10^{-2}$  & steady state \\
                                        & 20  &  $5.21\times10^{-8}$   &  $9.69\times10^{-3}$  & steady state \\
                                        & 50  &  $2.67\times10^{-8}$   &  $6.41\times10^{-3}$  & steady state \\
                                        & 100 &  $1.71\times10^{-8}$   &  $5.15\times10^{-3}$  & steady state \\
                                        & 200 &  $1.12\times10^{-8}$   &  $4.18\times10^{-3}$  & steady state \\
                                        & 500 &  $5.28\times10^{-9}$   &  $2.55\times10^{-3}$  & steady state \\ \hline
\multirow{7}{*}{$8\times10^{5}$~K}      & 5   &  $2.83\times10^{-7}$   &  $3.82\times10^{-2}$  & steady state \\
                                        & 10  &  $1.16\times10^{-7}$   &  $1.69\times10^{-2}$  & steady state \\
                                        & 20  &  $2.34\times10^{-8}$   &  $3.13\times10^{-3}$  & steady state \\
                                        & 50  &  $2.73\times10^{-10}$  &  $3.05\times10^{-5}$  & variable     \\
                                        & 100 &  $1.20\times10^{-11}$  &  $5.33\times10^{-7}$  & variable     \\
                                        & 200 &  $2.46\times10^{-13}$  &  $3.16\times10^{-9}$  & variable     \\
                                        & 500 &  $1.01\times10^{-11}$  &  $2.18\times10^{-6}$  & variable     \\ \hline
\multirow{7}{*}{$10^{6}$~K}             & 5   &  $2.22\times10^{-7}$   &  $3.10\times10^{-2}$  & steady state \\
                                        & 10  &  $1.03\times10^{-7}$   &  $1.54\times10^{-2}$  & steady state \\
                                        & 20  &  $3.05\times10^{-8}$   &  $4.52\times10^{-3}$  & steady state \\
                                        & 50  &  $4.62\times10^{-9}$   &  $9.84\times10^{-4}$  & steady state \\
                                        & 100 &  $3.31\times10^{-9}$   &  $1.38\times10^{-3}$  & variable     \\
                                        & 200 &  $3.20\times10^{-9}$   &  $2.43\times10^{-3}$  & variable     \\
                                        & 500 &  $2.96\times10^{-9}$   &  $4.54\times10^{-3}$  & variable     \\ \hline
\multirow{7}{*}{CAK}                    & 5   &  $4.91\times10^{-7}$   &  $5.26\times10^{-2}$  & steady state \\
                                        & 10  &  $4.70\times10^{-8}$   &  $6.58\times10^{-3}$  & steady state \\
                                        & 20  &  $2.36\times10^{-8}$   &  $6.23\times10^{-3}$  & steady state \\
                                        & 50  &  $2.09\times10^{-8}$   &  $9.60\times10^{-3}$  & steady state \\
                                        & 100 &  $2.05\times10^{-8}$   &  $1.34\times10^{-2}$  & steady state \\
                                        & 200 &  $2.03\times10^{-8}$   &  $1.88\times10^{-2}$  & steady state \\
                                        & 500 &  $2.01\times10^{-8}$   &  $2.94\times10^{-2}$  & steady state \\ \hline
\multirow{7}{*}{$\delta=0.1$}           & 5   &  $4.84\times10^{-7}$   &  $5.13\times10^{-2}$  & steady state \\
                                        & 10  &  $3.31\times10^{-8}$   &  $4.09\times10^{-3}$  & steady state \\
                                        & 20  &  $9.90\times10^{-9}$   &  $2.21\times10^{-3}$  & steady state \\
                                        & 50  &  $7.38\times10^{-9}$   &  $2.75\times10^{-3}$  & steady state \\
                                        & 100 &  $6.63\times10^{-9}$   &  $3.44\times10^{-3}$  & steady state \\
                                        & 200 &  $6.11\times10^{-9}$   &  $4.38\times10^{-3}$  & steady state \\
                                        & 500 &  $5.59\times10^{-9}$   &  $6.13\times10^{-3}$  & steady state \\ \hline
\multirow{7}{*}{$\delta=0.3$}           & 5   &  $4.73\times10^{-7}$   &  $4.93\times10^{-2}$  & steady state \\
                                        & 10  &  $1.91\times10^{-8}$   &  $1.97\times10^{-3}$  & steady state \\
                                        & 20  &  $7.33\times10^{-10}$  &  $1.20\times10^{-4}$  & steady state\\
                                        & 50  &  $2.60\times10^{-10}$  &  $7.33\times10^{-5}$  & variable    \\
                                        & 100 &  $1.79\times10^{-10}$  &  $7.07\times10^{-5}$  & variable    \\
                                        & 200 &  $1.20\times10^{-10}$  &  $6.59\times10^{-5}$  & variable    \\
                                        & 500 &  $9.14\times10^{-11}$  &  $7.67\times10^{-5}$  & variable    \\ \hline    
\end{tabular}
\caption{\footnotesize List of the simulations considered in this work and summary 
of some gross properties of the wind solutions. The first column list the model case, 
from the top to the bottom: self-consistent model with radiative heating and cooling 
and line driving based on the results from photoionization calculations 
for $4\times10^5$, $8\times10^5$, and $10^6$~BB SEDs, isothermal model using 
the \citetalias{CAK}, and two mCAK expressions for line driving with $\delta=0.1$ and 0.3.
The second column list $\hep0$ value. The third and fourth columns list
the wind mass loss rate and momentum efficiency, $\dot{M}_{w}$ and
$\eta_{wind}\equiv \dot{M}_{w} v_{out} c /(L_{\rm Edd}\Gamma)$, respectively.
We plot these two and other wind properties as functions of $\hep0$ in
Fig.~\ref{fig:HEP}. Finally, the last column provides information on whether 
the wind solution reached a steady state or remained variable.
}
\label{tab:result-sum}
\end{table*}

\quad

\subsection{Parameter selection}
The original application of the \citetalias{CAK} theory was to explain mass outflows from OB stars.
As we mentioned in \S\,\ref{sec:intro}, the theory has been applied to outflows in CVs 
and AGNs. In such accretion disk systems, the range of atmospheric temperatures is 
much wider than among OB stars; in CVs and AGNs, temperatures can be about $10^3$~K 
or less and as high as a few $10^5$~K typically. AGNs have the additional complication 
that the X-ray source compared to the thermal BB-like source 
can be significantly stronger than in CVs. 
To keep our exploration of the various radiative environments general 
and manageable, we will focus here on BB SED cases with 
the temperature significantly higher than that applicable to OB stars.

\begin{figure*}[htb!]
  \centering
  \includegraphics[width=0.8\textwidth]{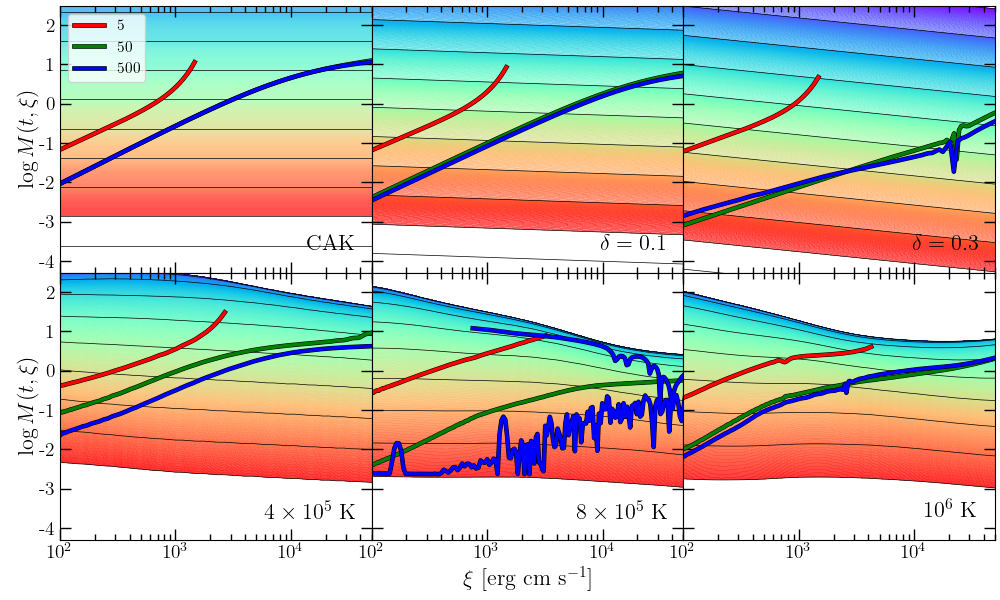}
  \caption{ 
  The force multiplier as a function of $\xi$ for a subset of our wind solutions. 
  We present six  cases: 
  1) CAK (i.e., $\delta=0$ top left panel); 
  2) mCAK with $\delta=0.1$ (top middle panel);
  3) mCAK with $\delta=0.3$ (top right panel);
   and 4-6) $4,\,8,\,10\,\times10^5$ BB SED cases (the bottom panels, from left to right).
   Each panel shows results of a single snapshot in time 
   for $\text{HEP}_0$=5, 50, and 500 (red, green, blue curves; 
   see Eq.~\ref{eq:hep}, for the definition of $\text{HEP}_0$). 
   For reference, in each panel, we also plot the force multiplier as a function of 
   $\xi$ for fixed values of optical depth parameter, $t$: $\log t$ = -8,-7,...,0,1).
   In the top panels, the solid lines are by design  straight lines, equally spaced 
   and with a slope equal to $-\delta$ (see the appendix including Eq.~\ref{eq:k_xi}). 
   In the bottom panels, the force multiplier is computed based on the photoionization 
   calculations and its $t$ and $\xi$ is more complex, for example, the solid lines converge 
   for small $t$ which reflects the saturation of \fmtxi as $t$ approaches zero. This saturation 
   is most apparent for the \eightbb~SED case and 
   has a significant effect on the wind solution especially for high $\hep0$ simulations 
   where the solution is highly time dependent. 
   Time dependent behaivor manifests on these figures as non-monotonic 
   behavior of $M$ as a function of $\xi$ (see Table~\ref{tab:result-sum}).  
   This is most dramatically shown in the panel corresponding to the \eightbb~SED case, 
   however one can see this effect in the panels for our $\delta$=0.3 and \milbb~SED cases as well.
   }
  \label{fig:xi}
\end{figure*}

In Table~\ref{tab:result-sum}, we summarize the main parameters for 
all simulations that we present in this paper. The table also lists 
the wind mass loss rate, $\dot{M}_{w}$, momentum efficiency 
$\eta_{wind}\equiv \dot{M}_{w} v_{out} c /(L_{\rm Edd}\Gamma)$ 
and whether the wind solution reached a steady state
or remained variable even after running the simulations for 2 or even 8
sound crossing times (i.e., a few 100 $r_0/c_{s,0}$).

Our illustrative wind simulations are 1D radial flows.  Therefore, it may come 
as a surprise that not all simulations resulted in steady-state solutions akin to 
time-independent \citetalias{CAK} solutions (see right column of Table~\ref{tab:result-sum}).   
We have investigated possible causes of this variable behavior and concluded 
it is not a numerical effect.  
So as to not limit the results of our parameter survey to steady solutions, 
these variable solutions have been included in our analysis, with wind properties 
quoted after time-averaging the solutions.  We plan on investigating in more detail 
the dynamics leading to variability in a future study. 
In \S\,\ref{sec:conclusions}, we offer a brief discussion of these variable solutions
and cite a couple of previous works that reported on oscillations in the outflows. 

\section{Results}
\label{sec:results}
Here, we  present our findings on wind solutions for different BB SEDs by focusing 
on three specific cases: BB SEDs with radiation temperatures $4\times10^5~{\rm K}$, $8\times10^5~{\rm K}$, and $10^6~{\rm K}$. 
We selected these cases based on the behavior of the force multiplier
(see Figs.~\ref{fig:fm}, \ref{fig:fm_a}, and \ref{fig:fm_p}).
The \fourbb~SED case is representative of the regime 
where the necessary condition for line driving (i.e., inequality~\ref{eq:mgone} graphically
shown as $M_{max}> 10$)
is satisfied for a wide range of $\xi$ whereas the \eightbb~SED and \milbb~SED cases 
are representatives of the regime where the necessary condition is satisfied only 
for a small or intermediate range of $\xi$. 
We view the \fourbb~SED  case as belonging to the class of cases similar to OB stars, 
where ionization effects are weak or nonexistent.
On the other hand, we categorize the latter two cases as belonging to the class of 
high-temperature BB SED cases, where ionization effects are strong.
However, we note that the ionization effects in high-temperature BB SED cases 
are not as strong as in AGN and XRB cases. Fig.~\ref{fig:fm} shows the force multiplier
and its characteristics for the three selected BB SED cases and for comparison for the AGN1 case. 
Figs.~\ref{fig:fm_a} and ~\ref{fig:fm_p} enable the comparison of the force multiplier 
in all 14 of our SED cases. 

As we mentioned in \S\ref{sec:method}, AGNs emit a significant amount of radiation 
in the form of ionizing photons that have the potential to completely ionize the gas 
and cause TI. This is even more true in XRBs. The two figures 
in Appendix~\ref{appendix-teq-v-fm} illustrate this point: 
$M_{max}$, $k$, and $\eta_{max}$ strongly decrease with increasing  $\xi$
for AGNs and XRBs,
which indicates a tendency towards complete ionization of the gas
(see corresponding panels in Fig. ~\ref{fig:fm_p}). 

In AGN and XRB cases, the slope of the S curve is greater than 1 for some values of $\xi$
indicating TI (refer to the solid black curve in the top right panel of Fig.~\ref{fig:fm} 
and in the panels corresponding to these SEDs in Fig.~\ref{fig:fm_a}).
In contrast, all our BB cases are thermally stable, as evident from the slope 
of the equilibrium curves in $\log T-\log \xi$ being less than 1 for all $\xi$ 
(see again the solid black curves in the top panels in Fig.~\ref{fig:fm}). Thus, we can expect that 
even for our highest $T_{\rm BB}$ cases, 
the wind solution will be qualitatively different from the thermally unstable AGN and XRB cases.

\subsection{Transition viewed as a function of $\xi$}
As we describe in Appendix~\ref{appendix-teq-v-fm}, $M_{max}$ typically decreases 
with increasing $\xi$ (see Figs.~\ref{fig:fm_a} and \ref{fig:fm_p}). 
This downward trend is especially strong for AGN 
and XRB SEDs and for BB SEDs with $T_{\rm BB} \gtrsim 4\times10^5$~K. 
However, for $T_{\rm BB} \lesssim 4\times10^5$~K, $M_{max}$
is relatively high and stays at about the level of 10
even for $\log \xi$  as high as 5. Thus, $T_{\rm BB}\approx 4-6\times10^5$~K marks 
the transition to the regime for which the line driving will not be significant 
because the necessary condition for driving is not satisfied for high $\xi$.

In Fig.~\ref{fig:xi}, we show how the force multiplier varies as a function of $\xi$ 
for 18 different wind solutions. We present results for $\text{HEP}_0$=~5,~50,~and~500 
(red, green, and blue curves) in six cases: 
1) CAK (i.e., $\delta=0$ top left panel); 
2) mCAK with $\delta=0.1$ (top middle panel);
3) mCAK with $\delta=0.3$ (top right panel);
and 4-6) $4\times 10^5~{\rm K}$, $8\times 10^5~{\rm K}$ and $10^6~{\rm K}$ BB cases (the bottom panels, from left to right).

The three CAK wind solutions correspond to the case we discuss in Appendix~\ref{appendix-relations}.
For $\hep0 \gtrsim 5$, when the gas is in the cold regime, 
the wind solution does not depend on $\hep0$  as predicted 
(see also Table~\ref{tab:input-sum} and Fig.~\ref{fig:HEP} where
we show some gross wind properties as functions of $\hep0$). For example, 
in the top left panel of Fig.~\ref{fig:xi}, the curves for $\hep0=\,$50 and 500
overlap and are different than the red curve corresponding to $\hep0=\,$5 which
is the hot solution where the wind is driven by the pressure force, not the line force.

In the $\delta=0.1$ mCAK case, we also see that the wind solutions for $\hep0$=50 
and 500 very similar. Fig.~\ref{fig:HEP} indicates that the transition between
the 'hot' to 'cold' solutions (i.e., to those  that become insensitive to $\hep0$) 
occurs at $\hep0 \approx 20$.

In the other four cases: mCAK with $\delta=0.3$, \fourbb, \eightbb, and \milbb~SED, 
the wind solutions depend on $\hep0$ for all $\hep0$.  These are the cases where 
the force multiplier quite strongly decreases  with increasing $\xi$ 
for a given $t$ (see, for instance, the slopes of the solid lines plotted 
in Fig.~\ref{fig:xi} and the decrease of $k$ in increasing $\xi$
in the lower panels of Fig.~\ref{fig:fm}). This decrease in the force multiplier 
with increasing $\xi$ results in the weakening of the wind as measured, for example by
the wind mass loss rate and velocity in units of the escape velocity that
both decrease with increasing $\hep0$ for large $\hep0$ 
(see the top and third from the top panel in Fig.~\ref{fig:HEP}).

\subsection{Transition viewed as a function of $\hep0$}
In Appendix~\ref{appendix-teq-v-fm}, we cited the expression for  
$\dot{M}_{w, {\rm CAK}}$ (see Eq.~\ref{eq:mdotcak}) and
derived an expression for the line optical depth parameter at the wind base,  
$t_0$ (i.e., Eq.~\ref{eq:t0}).
These two quantities are $\hep0$ independent for large $\hep0$,
under the assumption that $k$ is a constant. Our numerical solutions 
for the CAK case are consistent with this prediction as shown 
by the red lines in Fig.~\ref{fig:HEP}. In the other cases 
that we presented here, $k$ varies with $\xi$ either explicitly 
via a power-law scaling with $\xi$ or $v$ (i.e., non zero $\delta$ in mCAK cases) 
or implicitly when we use our photoionization calculations. In these cases, 
the wind solution depends on $\hep0$.

The main physical reason for this $\hep0$ dependency can be described 
in the following way: $\dot{M}_w$ and other key wind properties, 
including $t_0$ are determined at the wind critical radius 
\cite[$r_c$, see ][for an overview]{LC}. 
To gain insight, one can still refer to eqs. \ref{eq:mdotcak} and \ref{eq:t0} even 
if the $k$ parameter decreases as $\xi$ increases. However, then one must 
consider the actual value of $k$ at $r_c$.

In our survey, 
we increase $\hep0$ by decreasing $r_0$ which results in
an increase in the escape velocity, $v_{esc}$
and a decrease in the flow velocity at the base, $v_0$  
(see Table~\ref{tab:input-sum} and the second from left panel 
in Fig.~\ref{fig:radius} for specific examples). 
Thus, as $\hep0$ increases the wind velocity  (and also $\xi$) downstream
must increase by a large factor in order to reach $v_{esc}$, 
which means that $k$ at large radii is smaller than at $r_0$.
Consequently, to maintain the net outward force, $t$ must decrease 
and boost the force multiplier by a decrease
in the wind density or an increase in the velocity gradient or both.

To explicitly illustrate these dependencies, 
in Fig.~\ref{fig:radius}, we plot several wind properties and forces acting 
on the wind as functions of radius for the \milbb~SED case and three values 
of $\hep0$. The right panel shows explicitly that $\xi$ at the critical radius 
increases with increasing $\hep0$ (compare the position of the points marked 
by the black squares). The right-hand sides of  eqs.~\ref{eq:mdotcak} and \ref{eq:t0} 
depend therefore indirectly on $\hep0$ through $k$. Formally, there is also indirect 
dependence through $\alpha$. But $\alpha$ is a very weak function of $\xi$ 
even in our self-consistent cases (see the solid blue lines in Figs.~\ref{fig:fm} 
and \ref{fig:fm_a}).

Eqs.~\ref{eq:mdotcak} and \ref{eq:t0} are of less help to gauge the effects 
of changes in $M_{\max}$ or $\eta$ on the wind properties. But generally 
one can expect that too large  a decrease in $M_{max}$ with $\xi$ will lead 
to the line force's inability to remain stronger than gravity downstream 
\citep[see][for an earlier exploration of the effects of the force multiplier saturation 
on disk and wind solution using a modified CAK method]{Proga99b}. This limited growth
of the force multiplier can prevent the wind from developing, or from being strong
like found by \cite{Proga99b} and us here. But moreover the force multiplier saturation 
can lead to a highly variable wind solution as we found here in high $\hep0$ simulations 
for the \eightbb~SED cases (see, for instance, the blue line in the bottom middle panel 
of Fig.~\ref{fig:xi}).

\begin{figure}[htb!]
  \centering
  \includegraphics[width=0.5\textwidth]{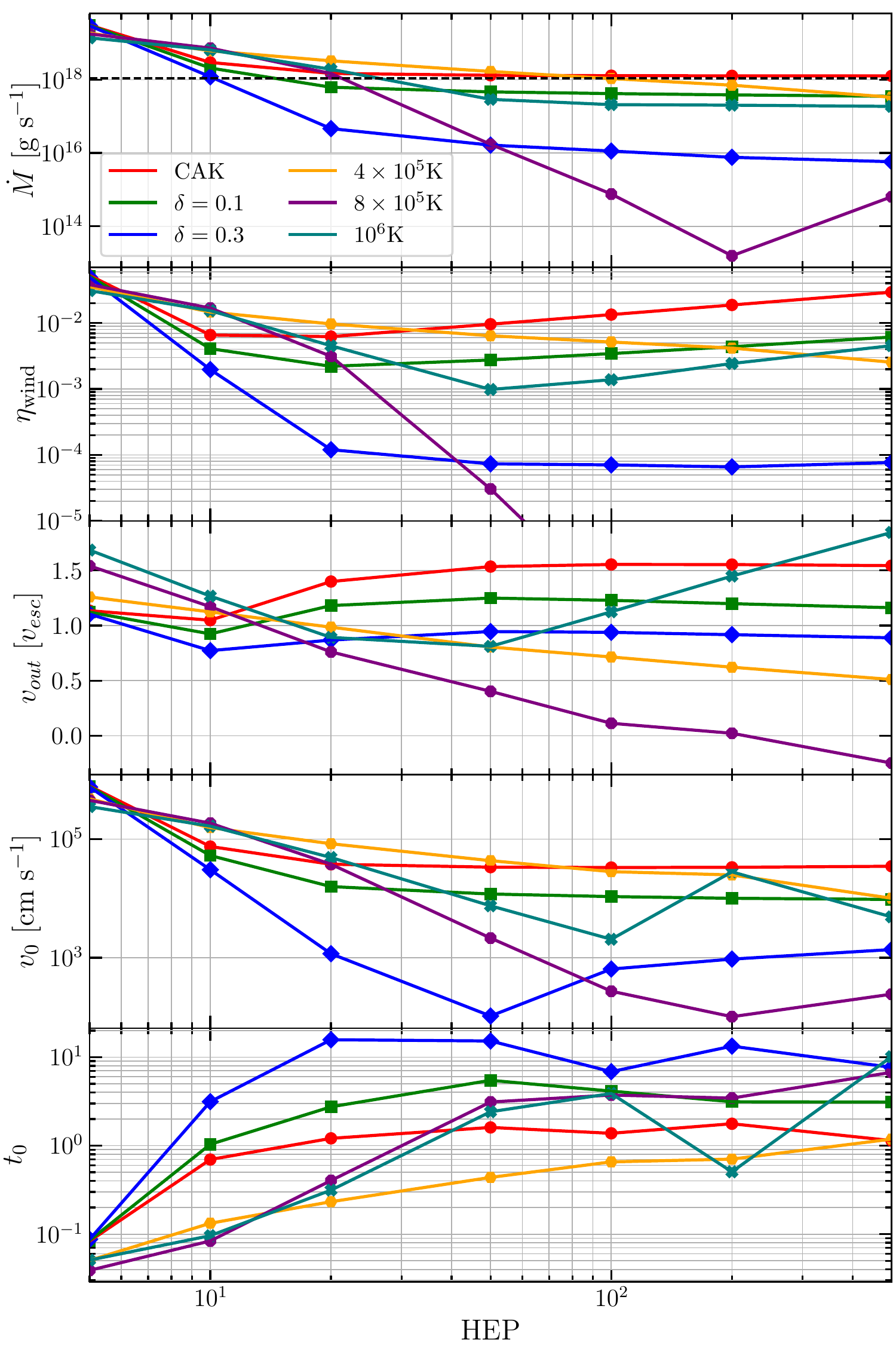}
  \caption{Comparison of the gross properties of our wind solutions  as a function 
  of $\hep0$.  From top to bottom, those quantities shown are the mass flux, 
  wind momentum efficiency $\eta_{\rm wind} = v\dot{M}\left(c/\left(\Gamma L_{\rm Edd}\right)\right)$,
  velocity as the gas exits the outer grid cell in units of escape velocity
  defined at the inner radius ($v_{esc} = \sqrt{2GM/r_{0}}$), $v_{out}$
  velocity at the inner radius, $v_0$,  
  and optical depth parameter at the inner radius, $t_0$ 
  The horizontal black dashed line in the top 
  panel represents $\dot{M}_{w, {\rm CAK}}$ (see Eq.~\ref{eq:mdotcak}).}
  \label{fig:HEP}
\end{figure}

\begin{figure*}[htb!]
  \centering
  \includegraphics[width=\textwidth]{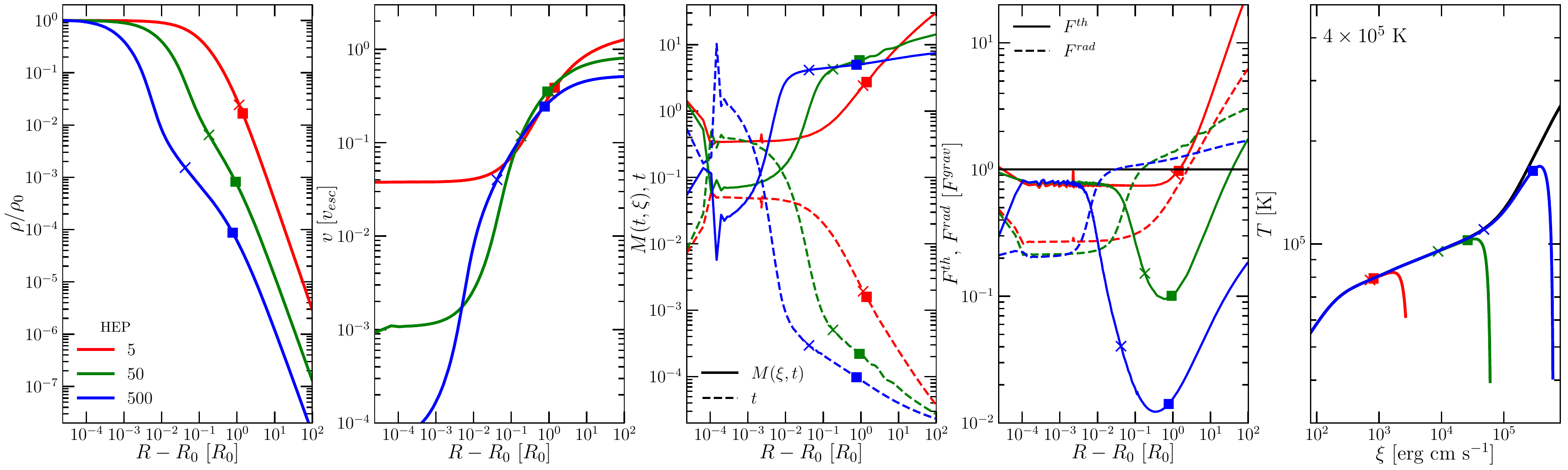}
  \caption{Properties of winds for the  \fourbb~SED case. We show three solutions 
  that correspond to $\hep0=5,50,500$ (red, green, and blue lines respectively).  
  Proceeding left to right, the first panel shows the density in units of the density 
  at $r_0$, $rho_0$. The second panel shows the radial velocity. The third panel shows 
  both the force multiplier, \fmtxi (solid), and the line optical depth parameter, $t$ (dashed), 
  as functions of radius.  The fourth panel displays the radiation force, $F^{rad}$, 
  and gas pressure, $F^{th}$, normalized to the force of gravity, $F^{grav}$, 
  as a function of radius. The final rightmost panel shows the phase diagram 
  (i.e., $\log T$ vs. $\log \xi$) for all three wind solutions. 
  The solid black curve corresponds to the thermal equilibrium ("S-curve", 
  i.e. where $T=T_eq(\xi)$). The x's and squares in matching colors on each curve mark 
  the sonic and critical radii, respectively).
   }
  \label{fig:radius}
\end{figure*}

\section{Discussion and Conclusions} \label{sec:conclusions}

AGNs emit radiation at a significant fraction of their corresponding Eddington luminosities 
\cite[and references therein]{Giustini19}. Therefore, the radiation force, including 
the line force 
must play a role in at least launching but even accelerating winds from accretion disks in these systems
(e.g., \citealp{ MCGV, PSK00, PK04, Nomura13, Nomura20}).
Such radiation-driven disk winds can avoid over-ionization because of self-shielding and
can account for the presence of broad absorption lines 
\citep[BALs; see][for a review of both observational and theoretical arguments]{Giustini19}. 

The robustness of the line launching was explored by \cite{PK04}
who considered the "worst case" scenario for line driving of disk winds in AGN.
Specifically, they showed that no self-shielding is needed to launch a wind from luminous
AGN where $\rm M_{\rm BH}$ is very high (i.e., $> 10^8~\MSUN$) 
so that the disk radiation is relatively soft and is dominated by UV radiation.
However, they assumed that the force multiplier parameters 
do not depend on the disk's local BB SED. This assumption was motivated by the results
from modeling of the line driving from OB stars, where it was shown that 
the energy distribution of the line opacity follows the SED as the star temperature changes 
and consequently $k$ and $\alpha$ do not significantly change with the SED 
\citep[e.g.,][]{Abbott82, LC}.
However, the temperature range in luminous  stars is relatively narrow, it ranges from 
~$10^4$~K to ~$5\times 10^4$~K. This range is certainly narrower than expected
in AGN disks where the inner disk temperature can be $10^5$~K if not enough much more.
Thus, it is possible that outside the OB temperature range the line force can significantly 
depend on the BB SED. In particulate, above some disk temperature, $k$ can decrease 
with increasing $T_{\rm BB}$.

It is in this context that we discuss our results here. We surveyed the parameter space of 
line forces resulting from various temperature BB SEDs, 
while also self-consistently calculating 
the heating and cooling rates to compute the temperature.
Our main finding is a quantification of how the line force and the wind properties depend 
on the assumed SED and the key parameters characterizing the wind base such $\hep0$.
Specifically, we found that the line force is relatively insensitive to the BB SEDs 
with a temperature as high as $2\times10^5$~K, so well above the temperature of OB stars. 
This means that the line force has a similar capability for launching wind over a wide range 
of disk radii including the radii close to the inner disk radius in very high $\rm M_{\rm BH}$
where much of the accretion luminosity is liberated. \footnote{This result also indicates 
that models of mass loss from the central objects of young planetary objects as well as 
accreting white dwarfs with the surface temperature of the order of $2\times10^5$~K could 
be viewed as a high-temperature extension of the models for OB stars.}

Our main conclusion is that the line force can operate over a very wide 
range of SEDs and of the ratio between the gravitational and thermal energies. 
In other words, this force can operate in both warm and cold regimes, and when 
the mean photon energy is relatively high, so well beyond the conditions for which 
the original CAK solutions were obtained, i.e., the cold regime of winds from OB stars.
Future modeling of the astrophysical outflows using
our method of computing self-consistently the radiative heating and cooling rate and
the line force will likely lead to new insights into the properties of line-driven winds, 
in particular such  basic ones like wind efficiency, velocity, and temporal behavior. 

\subsection{Intrinsically variable wind regime}
Here, we found that for some BB SEDs,  
as $\hep0$ increases, the wind weakens but can reach a steady state. 
However, there are also some BB SEDs (e.g., \eightbb~case) in which the wind weakens 
with increasing $\hep0$ but it does not reach a steady state. These variable
winds are in the cold regime, i.e., $\hep0 \gtrsim 50$. 

The gas in accretion disks in AGN, and also CVs is also in the cold regime.
These disks are geometrically thin because their thermal energy 
is small compared to the huge gravitational energy of the central compact object: 
the corresponding $\hep0$ are large ($>$ a few thousand even). 
\cite{Proga98, Proga99b} presented results axisymmetric time-dependent simulations of 
line-driven disk winds. They found a class of  wind solutions  that are intrinsically variable 
even though the force multiplier parameters were assumed to be constant, 
in particular $k=k_0$, which corresponds to $\delta=0$ 
(see also \citealt{Dyda20} for a confirmation of this result in the upgraded 
and 3-D counterparts of these simulations).
It is possible then that the process, responsible for wind variability in our simulations, 
could cause an increased variability in disk winds when $delta$ is non-zero, or generally, if 
the force multiplier parameters are allowed to depend on SED and $\xi$ as in our models.

In \S\ref{sec:results}, we noted that in the \eightbb~case, the $\delta$ parameter 
is relatively high, (i.e., $\sim0.3$, higher than for typical OB cases, \cite[see, e.g,][]{LC}.
To elucidate the cause of the variability in our solutions, we have explored 
a few isothermal cases using a modified CAK method (see Eq.~\ref{eq:mcak}). 
We found that even for these relatively simple cases
the wind is variable, even periodic, for $\delta=0.3$. Therefore, we attribute 
this variability to the weakening of the line force as the wind velocity increases.
(recall that $\delta$ is a measure how quickly $k$ decreases with increase $\xi$ 
or the wind velocity, see Eq.~\ref{eq:k_xi}).
We plan on further investigating this variable solution. However, two 
points are worth noting: 1) \cite{CCG2011} studied line-driven winds for various
$\delta$ and stated  that they were not able to find any steady-state wind solution 
in the interval $0.22 < \delta < 0.30$. We did find solutions in this interval
but they are unsteady; 2) in a stratified atmosphere, the velocity amplitude 
of the propagating sound waves increases with altitude 
(e.g., see \citealt{Clarke14} for a textbook elaboration on this point). 
While sound waves form at the base of all line driven winds, 
only when the line driving is efficiently accelerating the wind (small $\delta$) 
do these sound waves become stretched at large radii.  As opposed to winds 
where line driving is weak (large $\delta$), this stretching effect is much weaker and therefore, 
these waves persist even out to large radii.
The lower wind velocity 
which is related to the smaller acceleration appears to be the key: when the acceleration is 
large the waves are stretched downstream and do not affect the wind, while when 
the acceleration is small the wave survey over large distances, and the related pressure force 
can affect a wind that is driven by a weak line force. Our unsteady wind solutions could also 
be related to some previously reported radial cases. For example, \cite{Owocki94} found that the density 
at the wind base affects the wind solution including its temporal behavior. Specifically, 
a steady state transonic outflow exists for a narrow range of $\rho_0$. For too small $\rho_0$, 
the outflow is supersonic already at the base whereas for too large $\rho_0$. 
\cite{Owocki94} wrote that they "encountered a kind of boundary 'stiffness'
that  induces a persistent base oscillation in density and velocity."  Similar issues 
were noticed by \cite{Proga98} and \cite{Proga99a}. In addition, \cite{Proga99a} checked 
how line-driven disk wind solutions depend on the sound speed at the base, 
which is equivalent to the dependence on $\hep0$. He found that although  the mass loss rate 
and velocity do not depend on $c_s$, the time behavior of the isothermal winds does depend 
on it. For example, the fiducial state-state solution model from \cite{Proga98} recalculated 
with $c_s$ reduced by a factor of 3 (corresponding to an increase of $\hep0$ by a factor of 3)
is somewhat time-dependent: density  fluctuations originating in the wind base spread 
in the form of streams sweeping outward. Generally, he found that wind time dependence weakens
with increasing $c_s$ because the gas pressure effects get stronger with increasing $c_s$. 
Subsequently, higher gas pressure smooths the flow more effectively and in a larger region 
above the disk mid-plane as the size of the subcritical part of the  flow increases with $c_s$. 
Thus, in retrospect, our systematic parameter survey appears to include some cases
that have been explored before by others concerned with setting up 
the lower boundary conditions appropriate to simulate an outflow that is subsonic 
near the base, goes through a critical point, and becomes supersonic at some finite distance 
from the base.

\acknowledgments 
Support for this work was provided by the National Aeronautics and Space Administration 
under TCAN grant 80NSSC21K0496.

\appendix
\renewcommand{\theequation}{A.\arabic{equation}}
\setcounter{equation}{0}
\newcounter{dummy}
\setcounter{dummy}{\value{figure}}
\renewcommand{\thefigure}{A.\number\numexpr\value{figure}-\value{dummy}\relax}
\section{Dependence of the force multiplier on the SED and ionization parameter}\label{appendix-teq-v-fm}

\begin{figure*}[htb!]
  \centering
  \includegraphics[width=0.95\textwidth]{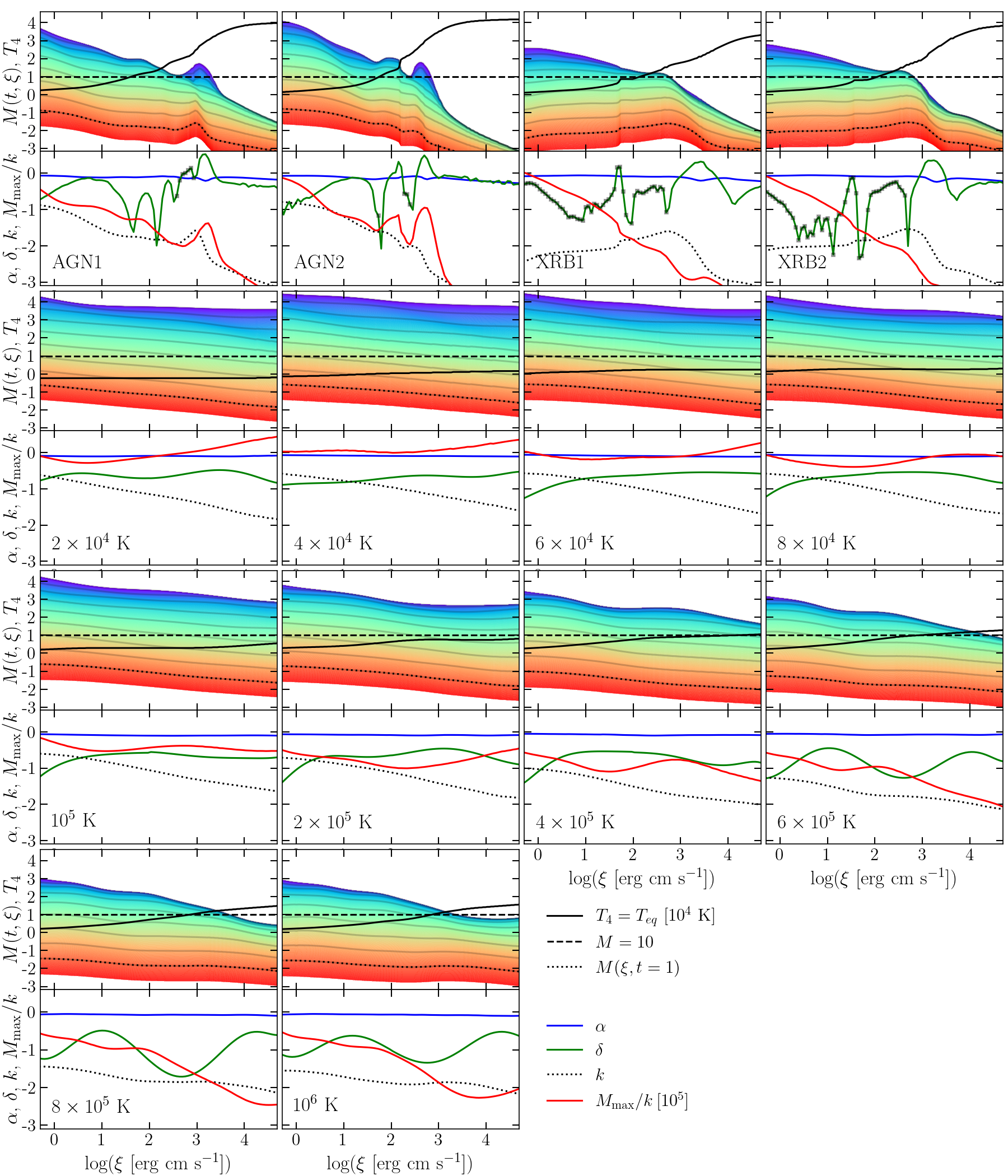}
  \caption{Parameter survey of photoionization results for the two AGNs and 
  two XRBs SEDs (top row of pair of panels), and ten blackbody SEDs of different temperatures.  
  From left to right,  from the second top row of panels to bottom, the SEDs correspond to BBs with 
  $T=2\times 10^{4}$~K,
  $4\times 10^{4}$~K,
  $6\times 10^{4}$~K,
  $8\times 10^{4}$~K,
  $10^{5}$~K,
  $2\times 10^{5}$~K,
  $4\times 10^{5}$~K,
  $6\times 10^{5}$~K,
  $8\times 10^{5}$~K,
  $10^{6}$~K.  
  The first four SEDs correspond to an unobscured and obscured AGN 
  \citep[AGN1 and AGN2;][]{Mehdipour15} and hard and soft state XRBs 
  \citep[XRB1 and XRB2;][]{Trigo13b}.
  {\textit{Top panels}}: colormap of the force multiplier for $t$ going 
  from $\log t = 1$ (red) to $\log t = -8$ (blue). Black dotted lines mark \fmtxi, 
  while gray lines indicate \fmtxi at $\log t= -1, -2, ... , -8$. 
  The horizontal dashed line marks the value \fmtxi$=10$, to indicate the level 
  above which line driving can be the dominant outflow driving mechanism. 
  The solid black line is the equilibrium temperature (or S-curve), $T_{\rm eq}$, 
  in units of $10^4$~K.
  {\textit{Bottom panels}}: profiles of $\alpha(\xi)$, $k(\xi)$, $|\delta(\xi)|$, 
  and $M_{max}(\xi)/k(\xi)$ (blue, black dashed, green, and red curves, respectively), 
  the four parameters characterizing the behavior of the force multiplier.To mark negative 
  values of $\delta$ we added gray squares atop the green line.
   }
  \label{fig:fm_a}
\end{figure*}

Here,  we summarize some results from our photoionization calculations of the force multiplier from \citetalias{Dyda17} and \citetalias{Dannen19}. 
Fig.~\ref{fig:fm_a} consists of fourteen pairs of panels for fourteen SEDs 
(as labeled in each lower panel, with `temperature labels' indicating the blackbody temeperature).
In the top panels, we plot the gas equilibrium temperature (solid black curve) as $T_{\rm eq}/10^4$~K.
The colored bands denote the force multiplier dependence on $\xi$;
the thick dotted line is for $M$ with $\log t=0$, while the other curves are, from the bottom to the top, with $\log t=-1, -2, ..., -8$.  The horizontal thick dashed line marks $M=10$. Our results show that regardless of type of the SED, for a fixed $\xi$, 
\fmtxi asymptotically approaches 0 as  $t$  becomes large 
and a constant maximum value, $M_{max}(\xi)$, as $t$ approaches zero; as a proxy for $M_{max}(\xi)$, we use $M(t=10^{-8},\xi)$. 
For intermediate values of $t$, the force multiplier  
monotonically increases with decreasing $t$. 

On the other hand, the dependence 
on $\xi$ for a fixed $t$ can be non-monotonic and specific to a given SED. 
For example, $M_{max}(\xi)$ is of order of a few $10^3$ for small 
$\xi$ in all SED cases, except for the two XRB cases where it is 
an order of magnitude smaller.  As $\xi$ increases, $M_{max}(\xi)$ decreases 
gradually for the BB SEDs with $T_{\rm BB} \lesssim 10^5$~K, strongly
for the hotter BB SEDs, and even more strongly for the AGN and XRBs cases.  This dependence is not part of the standard parameterization introduced by CAK, so here we analyze it in some detail.

\subsection{Ionization parameter dependence of the force multiplier I: CAK-like parameterization}
It might be possible to capture 
the force multiplier dependence on both $t$ and $\xi$ using some analytic 
formulae as, for example, \cite{SK90} did for a 10~keV bremsstrahlung.  
However, the fitting parameters would necessarily be specific to a given SED.
We opt to instead use the actual results from our photoionization
calculations in a tabulated form. Nevertheless, we find
that it is instructive and useful to relate 
our results to some well-known and often-used scaling relations.

The $t$-dependence of the force multiplier, including  its asymptotic behavior 
that we found are, unsurprisingly, consistent with the results first found by \citetalias{CAK}. 
They showed that for the large and intermediate values of $t$, the force multiplier 
can be well approximated by the following power law: 
\begin{equation}\label{eq:fmcak}
M_{\rm CAK}(t)=k_{\rm CAK} t^{-\alpha}, 
\end{equation}
where $k_{\rm CAK}$ and $\alpha$ are constants that can be estimated 
by fitting the above formula to the results from ionization calculations 
or using a theoretical  approach that considers a statistical distribution 
of the number of lines as a function of the line opacity and of the line frequency. 
The saturation of the force multiplier for very small $t$
corresponds to an optically thin limit for all lines,
including the strongest. They estimated  that $M_{max}$ is of the order of 
a few $10^3$ for OB stars. 

Motivated by the CAK parameterization, we express \fmtxi as
\begin{equation}\label{eq:fmc}
M(t, \xi)= M(t=1,\xi) t^{-\alpha} \epsilon_{fm} = k(\xi) t^{-\alpha} \epsilon_{fm},
\end{equation}
where $M(t=1,\xi)$ is a normalization factor, which is just a $\xi$-dependent version 
of  the $k_{CAK}$ parameter (hence the second equality) and 
$\epsilon_{fm} = \epsilon_{fm}(t,\xi)$ is a factor correcting the CAK-like scaling 
for additional dependence of \fmtxi on $t$ and $\xi$.  In particular, it accounts 
for the saturation of \fmtxi as $t$ approaches zero (see below for more discussion of this point).  

To parameterize the force multiplier dependence on ionization at moderate $t$, 
we compute the slope of the force multiplier at $t=1$:
$\delta(\xi) \equiv - \partial \log M(t,\xi)/\partial\log \xi |_{t=1}= d \log k(\xi)/ d \log \xi$. 
Here, we follow the approach adopted 
by \cite{Abbott82} who introduced this parameter to estimate the effect 
of ionization on the force multiplier through the factor 
$(n_{11}/W)^\delta$, where $n_{11}$ is the number density in units 
of $10^{11}~{\rm cm^3}~{\rm s}^{-1}$ and $W=W(r)$ is the geometrical dilution 
factor. In our photoionization calculations, $\delta$ corresponds to that 
introduced by \cite{Abbott82} because $J\propto W$ and using the definition
of $\xi$ (i.e., Eq.~\ref{eq:xi-def}), one finds that $\xi \propto W/n$.
We use $\delta$ to approximate the $k$-dependence of $\xi$:
\begin{equation}\label{eq:k_xi}
    k(\xi)\simeq k_0 (\xi/\xi_{0})^{-\delta}.
\end{equation}
Here, $k_0$ is the value of $k(\xi)$ for a fiducial value 
of the ionization parameter, $\xi_0$.

In the bottom panels of Fig.~\ref{fig:fm_a}, we plot four parameters characterizing 
the force multiplier using the convention from \citetalias{CAK} and \cite{Abbott82}. 
Specifically, we show
$\alpha(\xi) \equiv \partial \log M(t,\xi) / \partial \log t |_{t=1}$ (solid blue curve); 
$k(\xi) \equiv M(t=1,\xi)$ (black dashed curve; note, it is the same quantity shown 
as the thick  black dotted line in the corresponding top panel), 
$\delta$ (green curves or green curves with squares if  $\delta$ is negative, 
i.e., where $k$ increases with $\xi$) and finally 
$M_{max}(\xi)/k(\xi)~(= M(t=10^{-8},\xi)/M(t=1,\xi)$ in units of $10^5$, solid red curve).

The saturation parameter $\epsilon_{fm}(t,\xi)$ in \eqref{eq:fmc} accounts 
for an important modification of the original CAK parameterization (i.e., Eq.~\ref{eq:fmcak}) 
which does not have the property $M(t=0) = M_{max}$. \cite{Owocki88} first remedied this by 
modifying the CAK statistical model through the introduction of a cut off in the maximum 
line strength, thereby limiting the effect of very strong lines.  
Specifically, they used the following expression:
\begin{equation}\label{eq:ocr}
\epsilon_{fm}(t) =
\left[ \frac{(1+t \eta_{max})^{1-\alpha}-1}
{(t \eta_{max})^{1-\alpha}}\right]= 
\left[ \frac{(1+\tau_{max})^{1-\alpha}-1}
{(\tau_{max})^{1-\alpha}}\right],
\end{equation}
where $\eta_{max}$ is the opacity of the most opaque line 
and $\tau_{max}=t\eta_{max}$ ($\eta_{max}$ is in units of electron scattering opacity). 
For very large $\tau_{max}$, one recovers the original CAK relation since $\epsilon_{fm} = 1$, 
whereas for small $\tau_{max}$, $\epsilon_{fm} = (1-\alpha)\tau_{max}^\alpha$ so that the force multiplier
becomes $M_{max}=k_{\rm CAK}(1-\alpha)\eta_{max}^\alpha$.  Solving this expression for $\eta_{max}$ gives 
$\eta_{max} = [M_{max}/ k / ( 1-\alpha )]^ {1/\alpha}$, and this expression provides an estimate of the cut-off opacity given the $\xi$-dependent parameters $M_{max}$, $k$, and $\alpha$ shown in Fig.~\ref{fig:fm_a}:
\begin{equation} \label{eq:etamax}
\eta_{max}(\xi) \simeq [M_{max}(\xi)/ k(\xi) / ( 1-\alpha(\xi) )]^ {1/\alpha(\xi)}.
\end{equation}
Hence, in general $\epsilon_{fm}$ can be considered a function of both $t$ and $\xi$.

\subsection{Ionization parameter dependence of the force multiplier II: Analysis}  
The four parameters $\alpha(\xi)$, $\delta(\xi)$, $k(\xi)$, and $M_{max}(\xi)$ can be used 
to evaluate the deviation of the force multiplier shown in the top panels of Fig.~\ref{fig:fm_a} 
from the scaling introduced by \citetalias{CAK}. In the CAK scaling, $\alpha$ and $k$ 
are independent of $\xi$, $M_{max}$ is formally infinite, and $\delta$ equals 0.  
For reference, it should be noted that under CAK's scaling,
the \fmtxi lines bracketing the colored bands would be purely horizontal 
and equally spaced from one another  
(as shown in the top-left panel in Fig.~\ref{fig:xi}).\par

Our results show that $\alpha \approx 0.8-0.9$ for all BB SEDs, suggesting that 
it is a relatively very weak function of $\xi$. 
However, for AGN1/2 and XRB1/2 SEDs, $\alpha$ exhibits a somewhat stronger 
dependence on $\xi$ and shows a slight decreasing trend with increasing $\xi$. Specifically, 
it decreases from approximately $0.8-0.9$ 
for $\xi$ values below or equal to $\log \xi \approx 2$ to around 
$0.5-0.6$ for $\log \xi=5$.

The other quantities are more sensitive to $\xi$ especially
for high-temperature BBs and AGN and XRB cases.
This was also found by \cite{SK90} for the 10~keV bremsstrahlung.
The dependence of the force multiplier on $\xi$ 
is reflected in a non-zero value of $\delta$ which could be
as high 1 in the cases with relatively hard SEDs. 
This value of $\delta$ is much higher than in OB stars where $\delta<0.2$
\citep[see for example Table~8.2 in][and references therein]{LC}
and in our results for BBs where $\delta$ does not exceed  0.4.
We note that $\delta$ is negative for some $\xi$, which
is an indication that the force multiplier can be a non-monotonic function of $\xi$
as found by \citetalias{Dannen19}. 
\citetalias{CAK} showed $k$ is related to the total number of spectral
lines. Therefore, the dependence of $k$ on $\xi$ is related to 
a variation in total line number, so it is not too surprising that 
$\delta$ could be negative for some $\xi$ \cite[note that a possibility 
of $\delta$ being negative in OB star cases has been mentioned by] []{Puls00}. 

The dependence of $k$ on $\xi$ is dynamically important because,
for example, the wind mass loss rate depends on 
$k$ (see Eq.~\ref{eq:mdotcak}). 
A visual inspection of our results indicates that the $k-$ and $\delta-\xi$ dependencies 
can be well captured by a single power law for BBs with temperatures 
$\lesssim 10^5$~K, while a broken-power-law would be 
required for higher temperature BBs. The AGN and XRBs cases are 
more complicated, as both $k$ and $\delta$ are non-monotonic functions 
of $\xi$. Piecewise functions (e.g., a combination of several power laws 
or power laws and exponential functions) would be required, 
as also found by \citet{SK90}.

In our hydrodynamical simulations, we have used tabulated values of \fmtxi~
rather than fits to these values. Nevertheless, the $\delta$ parameterization
of the $k-\xi$ dependence (see Eq.~\ref{eq:k_xi}) has proven to be
very useful in interpreting our hydrodynamical results and making
a more direct connection with previous work by \cite{Abbott82} and others.

\begin{figure}[htb!]
  \centering
  \includegraphics[width=0.92\textwidth]{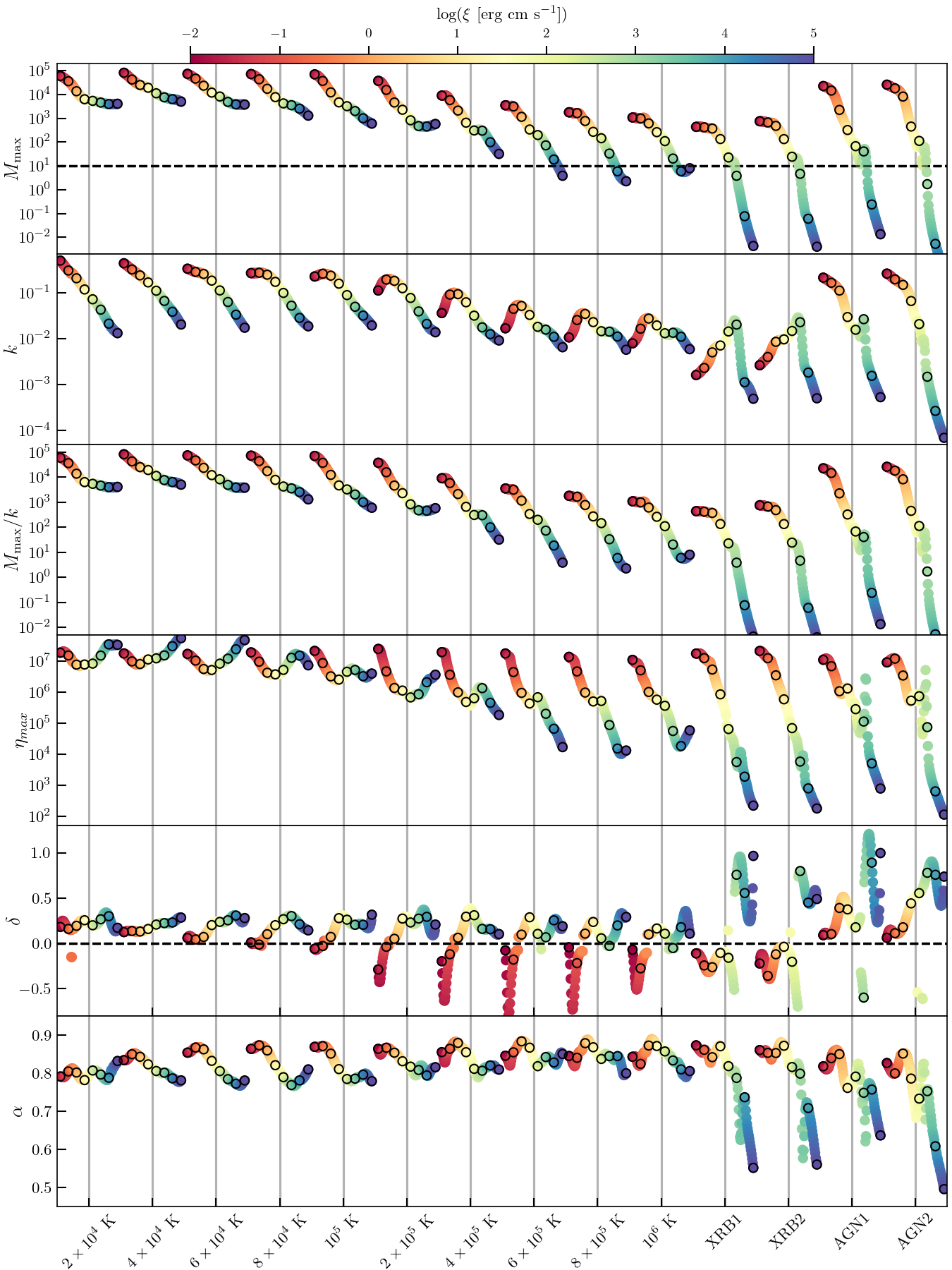}
  \caption{Parameter survey of our line force results for the ten BB SEDs 
  of different temperatures,  XRB, and AGN SEDs with increasing mean photon energy 
  from left to right.  The parameters of interest from top to bottom are $M_{max}$, $k$, 
  the ratio of $M_{max}$ and $k$, maximum line opacity $\eta_{max}$, $\alpha$, and $\delta$.  
  The color of each point corresponds to $\log\xi$ shown in the color bar along 
  the top (red corresponding to low ionization and blue with high ionization).  
  The circled points correspond to where $\log\xi = -2,-1,0,1,2,3,4,5$, 
  and the vertical lines are for reference but also indicate where $\xi\approx30$ 
  for each SED.}
  \label{fig:fm_p}
\end{figure}

To facilitate a comparison of the properties of the force multiplier 
for different SEDs and their trends with $\xi$, we show six parameters
in Fig.~\ref{fig:fm_p}:
$M_{max}$, $k$, the ratio of $M_{max}$ and $k$, maximum line opacity $\eta_{amx}$, 
$\alpha$, and $\delta$. One of the main results shown in this figure is that
up to $T_{\rm BB} \sim 2\times 10^5$~K, the properties of \fmtxi are not very sensitive to SEDs.
For higher temperature SEDs, the force multiplier parameters are sensitive to SEDs and 
strongly dependent on $\xi$, except for the $\alpha$ parameter 
which only varies by a factor of 2 across a range of seven orders of magnitude of $\xi$, 
even for the hottest SED.

\section{Scaling relations} \label{appendix-relations}
\renewcommand{\theequation}{B.\arabic{equation}}
A useful measure of the strength of the thermally driven wind 
is the ratio of the effective gravitational potential 
to the thermal energy at the radius of the wind base, $r_0$, 
usually termed the hydrodynamic escape parameter (HEP), that is,
\begin{equation} \label{eq:hep}
\hep0 = \frac{G M (1 - \Gamma)}{r_0 c_{s}^{2}},
\end{equation}
where $\rm M$ is the central mass, $c_s$ the sound speed, 
and $\Gamma$ is the luminosity in units of the Eddington luminosity $L_{Edd}$. 
For $\hep0 \lesssim 10$, a thermally driven hydrodynamic (Parker) wind will 
be produced \citep[e.g.,][and reference therein]{Stone09}. 
This parameter is related to the density scale height 
of an isothermal static atmosphere, $\lambda_\rho\equiv|\rho/(d \rho/d r)|=r_0/\hep0$.
At the base of an outflowing atmosphere, the gas velocity is
very sensitive to $\hep0$ for $\hep0>$ a few,
[i.e., $v_{0,{\rm th}} \simeq c_s (\hep0/2)^2 \exp{(-\hep0+3/2})$, see e.g., 
\citet{LC}].
Thus the mass loss rate of the thermally driven wind also exponentially decreases 
with increasing $\hep0$ (i.e., 
$\dot{M}_{w, th} \simeq 4\pi r_0^2 \rho_0 v_0$, where $\rho_0$ is the gas 
density at the wind base.) This dimensionless parameter 
is also useful in measuring the strength of the thermal wind 
from an irradiated disk \citep[see, e.g.,][]{Waters21}.

For line driving, such a measure of the wind strength should also exist but
determining it is somewhat more involved.
The basic requirement for line driving to dominate gravity is 
\begin{equation} \label{eq:mgone}
M_{\rm max} \Gamma \gtrsim 1,
\end{equation}
where $M_{\rm max}$ is the maximum value of the force multiplier 
for a given $\xi$ when even most opaques lines are optically thin. 
The value of $M_{max}$ could be as high as a few  $\times 10^3$ for low $\xi$, 
but it depends on $\xi$ and can be viewed as the upper limit of \fmtxi when 
$t$ approaches zero.
However, the above requirement is just a necessary and not a sufficient 
condition for producing an appreciable line-driven wind, because the actual value 
of $M$ can be smaller than $M_{max}$ as the optical parameter is not always 
very much smaller than unity. Using the Sobolev approximation, 
\citet[][\citetalias{CAK} hereafter]{CAK}
showed that $t\propto \rho /(dv_l/d l)$.
Thus more information is needed to compute the actual value of $M$ than to compute $M_{max}$.

To estimate $t$, we will use the analytic expression for the mass-lose of the line-driven
wind found by \citetalias{CAK}: 
\begin{equation} \label{eq:mdotcak}
\dot{M}_{w, {\rm CAK}} = \frac{4\pi G M(1 - \Gamma)}{\sigma_e \vth} 
\left[\frac{\alpha}{1-\alpha} \left(k_{\rm CAK}\Gamma \frac{1-\alpha}{1-\Gamma} \right)^{1/\alpha} \right].
\end{equation}
Next, we estimate the gas properties near the wind base
where gas is gravitationally bound and subsonic
so that the density radial profile is very close 
to the profile of an atmosphere in hydrostatic equilibrium 
($1/\lambda_\rho= -\hep0/r_0$).
On the other hand, using the mass continuity  
equation for a steady state, isothermal, spherical wind
we have $1/\lambda_\rho + 1/\lambda_v +2/r=0$
where $\lambda_v=v/(dv/dr)$ is the velocity scale length
which at the wind base can be approximated
as $\lambda_{v,0} \simeq r_0/(\hep0 -2)$.
Finally, using the last expression and the expressions 
for $\dot{M}_{w, {\rm CAK}}$ and $t$, we estimate
the optical depth parameter at the wind base as
\begin{equation} \label{eq:t0HEP}
t_0 \simeq \tau_0^2 \frac{v_{th}^2}{c_s^2}
\frac{1}{\hep0 (\hep0-2)}
         \frac{1-\alpha}{\alpha}\left[\frac{\Gamma}{1-\Gamma} k (1 - \alpha) \right]^{-1/\alpha},
\end{equation}
where $\tau_0=\sigma_e \rho_0 r_0$.  
The $\hep0$ dependence in Eq.~\ref{eq:t0HEP} can be eliminated when we consider cases 
for  $\hep0> 2$ and use the definitions of $\xi$ and $\hep0$.
Namely, using Eqs.~\ref{eq:xi-def} and \ref{eq:hep}, we can write
\begin{equation} \label{eq:tau0xi}
\tau_0 = \frac{4 \pi c k_B T_{\rm eq}(\xi_0)}{\xi_0} \frac{\Gamma}{1-\Gamma} \hep0
\end{equation}
or introducing the so-called pressure ionization parameter, 
\begin{equation}\label{eq:Xi-def}
    \Xi=\xi/(4\pi c k_B T),
\end{equation}
we can rewrite it as
\begin{equation} \label{eq:tau0}
\tau_0 = \frac{\Gamma}{1-\Gamma} \frac{\hep0}{\Xi_0},
\end{equation}
to find  that
\begin{equation} \label{eq:t0}
t_{0} \simeq  \left( \frac{\Gamma}{1 - \Gamma}\right)^{2}
        \frac{1-\alpha}{\alpha}\left[\frac{\Gamma}{1-\Gamma} k (1 - \alpha) \right]^{-1/\alpha} \left(\frac{v_{th}}{c_s}\right)^2\Xi_0^{-2}.
\end{equation}
Note that the sound speed is a function of $\xi_0$ via the gas temperature 
(i.e., for thermal equilibrium 
$\mathcal{L}(\xi_{0},T_{\rm eq,0})=0$),
whereas $v_{th}$ is kept constant in calculations of $k$ and $\alpha$
\citep[e.g.,][]{SK90}.

In \citetalias{Dannen20}, we found that for given black hole mass $M_{\rm BM}$ and SED, 
three parameters govern the solutions: $\Gamma$, $\Xi_0$, and $\hep0$.
These parameters set the strength of thermal driving, for example 
the mass loss of the thermal wind can be estimated as
\begin{equation} \label{eq:mdotthXi}
\dot{M}_{w, th}=
\frac{GM\Gamma}{4 c_s \sigma_e}
\frac{\hep0^2}{\exp(\hep0-3/2)}{\Xi_0}^{-1}.
\end{equation}
We could also estimate the optical depth parameter, $t$, at the base of 
a thermal wind, $t_{0, th}$.
Using the same approach that we used to estimate $t_0$ and expression for
$\dot{M}_{w, th}$, we find 
\begin{equation}\label{eq:t0th}
t_{0, th} = 4 \frac{1-\Gamma}{\Gamma}
        \frac{\exp(\hep0-3/2)}{\hep0(\hep0-2)} \frac{v_{th}}{c_s}\Xi_0^{-1} .
\end{equation}

Here, we show that two of these three dimensionless parameters, i.e.,
$\Gamma$ and $\Xi_0$, set also the strength of line driving, for given
$k$ and $\alpha$. Eq.~\ref{eq:t0} can be used to explain why line driving 
was negligible in the AGN cases we explored earlier.
For the wind base to have a relatively low temperature so that 
the AGN-irradiated gas 
could be on a cold stable branch of the thermal equilibrium 
and to have many spectral lines,  $\Xi_0$ needs to be less than 10 
\citepalias[e.g., see Fig.~1 in][]{Dannen20}.
However, this does not lead to a large force multiplier because $t$
at the base, $t_0$ is not a free parameter: it decreases with increasing $\Xi_0$ 
and therefore, it is large for small $\Xi_0$. 

The line optical depth dependence on $\Xi_0$ determines the parameter space 
for line driving in our self-consistent model: if $\Xi_0$ is low enough 
for many lines to exist (i.e., large $k$ and $M_{max}$), the line optical depth
could be too large for the line force to operate. To reduce the line's optical depth,
$\Xi_0$ would need to be increased but this could reduce the number 
of driving lines (i.e., small $k$ and $M_{max}$). 
Equation~\ref{eq:t0th} shows that $t$ at the base at a thermally driven wind 
exponentially increases with $\hep0$ for large $\hep0$ which implies that 
as the thermal wind weakens with increasing $\hep0$, the line driving 
may not necessarily kick in and strengthen the wind because the gas could 
be too optically thick unless $\Xi$ or $c_s$ or both increase.

In the two AGN cases explored in \citetalias{Dannen20}, we found that there is 
no parameter space for which \fmtxi$\Gamma>$ exceeds 1. 
This is despite the fact that for a wide range of $\xi$ values, 
the necessary condition for line driving is satisfied.
In section \S\,\ref{sec:results}, we present specific examples of BB SED cases 
that meet not only the necessary condition stated in inequality \ref{eq:mgone}
but also the \fmtxi$\Gamma>1$ condition.
These cases can have a BB temperature, $T_{\rm BB}$, as high as $10^6$~K.


\begin{thebibliography}{}
\expandafter\ifx\csname natexlab\endcsname\relax\def\natexlab#1{#1}\fi
\providecommand{\url}[1]{\href{#1}{#1}}
\providecommand{\dodoi}[1]{doi:~\href{http://doi.org/#1}{\nolinkurl{#1}}}
\providecommand{\doeprint}[1]{\href{http://ascl.net/#1}{\nolinkurl{http://ascl.net/#1}}}
\providecommand{\doarXiv}[1]{\href{https://arxiv.org/abs/#1}{\nolinkurl{https://arxiv.org/abs/#1}}}

\bibitem[{{Abbott}(1982)}]{Abbott82}
{Abbott}, D.~C. 1982, \apj, 259, 282, \dodoi{10.1086/160166}

\bibitem[{{Arav} \& {Li}(1994)}]{Arav94}
{Arav}, N., \& {Li}, Z.-Y. 1994, \apj, 427, 700, \dodoi{10.1086/174177}

\bibitem[{{Blandford} \& {Payne}(1982)}]{BlandfordPayne}
{Blandford}, R.~D., \& {Payne}, D.~G. 1982, \mnras, 199, 883,
  \dodoi{10.1093/mnras/199.4.883}

\bibitem[{{Bowler} {et~al.}(2014){Bowler}, {Hewett}, {Allen}, \&
  {Ferland}}]{Bowler14}
{Bowler}, R. A.~A., {Hewett}, P.~C., {Allen}, J.~T., \& {Ferland}, G.~J. 2014,
  \mnras, 445, 359, \dodoi{10.1093/mnras/stu1730}

\bibitem[{{Castor}(2007)}]{Castor07}
{Castor}, J.~I. 2007, {Radiation Hydrodynamics}

\bibitem[{{Castor} {et~al.}(1975){Castor}, {Abbott}, \& {Klein}}]{CAK}
{Castor}, J.~I., {Abbott}, D.~C., \& {Klein}, R.~I. 1975, \apj, 195, 157,
  \dodoi{10.1086/153315}

\bibitem[{{Clarke} \& {Carswell}(2014)}]{Clarke14}
{Clarke}, C., \& {Carswell}, B. 2014, {Principles of Astrophysical Fluid
  Dynamics}

\bibitem[{{Cur{\'e}} {et~al.}(2011){Cur{\'e}}, {Cidale}, \&
  {Granada}}]{CCG2011}
{Cur{\'e}}, M., {Cidale}, L., \& {Granada}, A. 2011, \apj, 737, 18,
  \dodoi{10.1088/0004-637X/737/1/18}

\bibitem[{{Dannen} {et~al.}(2019){Dannen}, {Proga}, {Kallman}, \&
  {Waters}}]{Dannen19}
{Dannen}, R.~C., {Proga}, D., {Kallman}, T.~R., \& {Waters}, T. 2019, \apj,
  882, 99, \dodoi{10.3847/1538-4357/ab340b}

\bibitem[{{Dannen} {et~al.}(2020){Dannen}, {Proga}, {Waters}, \&
  {Dyda}}]{Dannen20}
{Dannen}, R.~C., {Proga}, D., {Waters}, T., \& {Dyda}, S. 2020, \apjl, 893,
  L34, \dodoi{10.3847/2041-8213/ab87a5}

\bibitem[{{de Kool} \& {Begelman}(1995)}]{dKB95}
{de Kool}, M., \& {Begelman}, M.~C. 1995, \apj, 455, 448,
  \dodoi{10.1086/176594}

\bibitem[{{Dyda} {et~al.}(2017){Dyda}, {Dannen}, {Waters}, \& {Proga}}]{Dyda17}
{Dyda}, S., {Dannen}, R., {Waters}, T., \& {Proga}, D. 2017, \mnras, 467, 4161,
  \dodoi{10.1093/mnras/stx406}

\bibitem[{{Dyda} {et~al.}(2020){Dyda}, {Proga}, \& {Reynolds}}]{Dyda20}
{Dyda}, S., {Proga}, D., \& {Reynolds}, C.~S. 2020, \mnras, 493, 437,
  \dodoi{10.1093/mnras/staa304}

\bibitem[{{Field}(1965)}]{Field65}
{Field}, G.~B. 1965, \apj, 142, 531, \dodoi{10.1086/148317}

\bibitem[{{Foltz} {et~al.}(1987){Foltz}, {Weymann}, {Morris}, \&
  {Turnshek}}]{Foltz87}
{Foltz}, C.~B., {Weymann}, R.~J., {Morris}, S.~L., \& {Turnshek}, D.~A. 1987,
  \apj, 317, 450, \dodoi{10.1086/165290}

\bibitem[{{Frank} {et~al.}(2002){Frank}, {King}, \& {Raine}}]{Frank02}
{Frank}, J., {King}, A., \& {Raine}, D.~J. 2002, {Accretion Power in
  Astrophysics: Third Edition}

\bibitem[{{Ganguly} {et~al.}(2003){Ganguly}, {Masiero}, {Charlton}, \&
  {Sembach}}]{Ganguly03}
{Ganguly}, R., {Masiero}, J., {Charlton}, J.~C., \& {Sembach}, K.~R. 2003,
  \apj, 598, 922, \dodoi{10.1086/379057}

\bibitem[{{Giustini} \& {Proga}(2019)}]{Giustini19}
{Giustini}, M., \& {Proga}, D. 2019, \aap, 630, A94,
  \dodoi{10.1051/0004-6361/201833810}

\bibitem[{{Gupta} {et~al.}(2003){Gupta}, {Srianand}, {Petitjean}, \&
  {Ledoux}}]{Gupta03}
{Gupta}, N., {Srianand}, R., {Petitjean}, P., \& {Ledoux}, C. 2003, \aap, 406,
  65, \dodoi{10.1051/0004-6361:20030793}

\bibitem[{{Higginbottom} {et~al.}(2014){Higginbottom}, {Proga}, {Knigge},
  {Long}, {Matthews}, \& {Sim}}]{Higginbottom14}
{Higginbottom}, N., {Proga}, D., {Knigge}, C., {et~al.} 2014, \apj, 789, 19,
  \dodoi{10.1088/0004-637X/789/1/19}

\bibitem[{{Jiang}(2022)}]{Jiang22}
{Jiang}, Y.-F. 2022, \apjs, 263, 4, \dodoi{10.3847/1538-4365/ac9231}

\bibitem[{{Kallman} \& {Bautista}(2001)}]{KB01}
{Kallman}, T., \& {Bautista}, M. 2001, \apjs, 133, 221, \dodoi{10.1086/319184}

\bibitem[{{Krolik}(1999)}]{Krolik99}
{Krolik}, J.~H. 1999, {Active galactic nuclei : from the central black hole to
  the galactic environment}

\bibitem[{{Kurosawa} \& {Proga}(2009)}]{Kurosawa09}
{Kurosawa}, R., \& {Proga}, D. 2009, \apj, 693, 1929,
  \dodoi{10.1088/0004-637X/693/2/1929}

\bibitem[{{Lamers} \& {Cassinelli}(1999)}]{LC}
{Lamers}, H. J.~G.~L.~M., \& {Cassinelli}, J.~P. 1999, {Introduction to Stellar
  Winds}

\bibitem[{{Lu} \& {Lin}(2018)}]{Lu18}
{Lu}, W.-J., \& {Lin}, Y.-R. 2018, \apj, 863, 186,
  \dodoi{10.3847/1538-4357/aad411}

\bibitem[{{Mas-Ribas} \& {Mauland}(2019)}]{MasRibas19}
{Mas-Ribas}, L., \& {Mauland}, R. 2019, \apj, 886, 151,
  \dodoi{10.3847/1538-4357/ab4efd}

\bibitem[{{Mehdipour} {et~al.}(2015){Mehdipour}, {Kaastra}, {Kriss}, {Cappi},
  {Petrucci}, {Steenbrugge}, {Arav}, {Behar}, {Bianchi}, {Boissay},
  {Branduardi-Raymont}, {Costantini}, {Ebrero}, {Di Gesu}, {Harrison}, {Kaspi},
  {De Marco}, {Matt}, {Paltani}, {Peterson}, {Ponti}, {Pozo Nu{\~n}ez}, {De
  Rosa}, {Ursini}, {de Vries}, {Walton}, \& {Whewell}}]{Mehdipour15}
{Mehdipour}, M., {Kaastra}, J.~S., {Kriss}, G.~A., {et~al.} 2015, \aap, 575,
  A22, \dodoi{10.1051/0004-6361/201425373}

\bibitem[{{Mihalas} \& {Mihalas}(1984)}]{MihalasMihalas84}
{Mihalas}, D., \& {Mihalas}, B.~W. 1984, {Foundations of radiation
  hydrodynamics}

\bibitem[{{Murray} {et~al.}(1995){Murray}, {Chiang}, {Grossman}, \&
  {Voit}}]{MCGV}
{Murray}, N., {Chiang}, J., {Grossman}, S.~A., \& {Voit}, G.~M. 1995, \apj,
  451, 498, \dodoi{10.1086/176238}

\bibitem[{{Mushotzky} {et~al.}(1972){Mushotzky}, {Solomon}, \&
  {Strittmatter}}]{Mushotzky72}
{Mushotzky}, R.~F., {Solomon}, P.~M., \& {Strittmatter}, P.~A. 1972, \apj, 174,
  7, \dodoi{10.1086/151463}

\bibitem[{{Nomura} {et~al.}(2020){Nomura}, {Ohsuga}, \& {Done}}]{Nomura20}
{Nomura}, M., {Ohsuga}, K., \& {Done}, C. 2020, \mnras, 494, 3616,
  \dodoi{10.1093/mnras/staa948}

\bibitem[{{Nomura} {et~al.}(2013){Nomura}, {Ohsuga}, {Wada}, {Susa}, \&
  {Misawa}}]{Nomura13}
{Nomura}, M., {Ohsuga}, K., {Wada}, K., {Susa}, H., \& {Misawa}, T. 2013,
  \pasj, 65, 40, \dodoi{10.1093/pasj/65.2.40}

\bibitem[{{North} {et~al.}(2006){North}, {Knigge}, \& {Goad}}]{North06}
{North}, M., {Knigge}, C., \& {Goad}, M. 2006, \mnras, 365, 1057,
  \dodoi{10.1111/j.1365-2966.2005.09828.x}

\bibitem[{{Owocki} {et~al.}(1988){Owocki}, {Castor}, \& {Rybicki}}]{Owocki88}
{Owocki}, S.~P., {Castor}, J.~I., \& {Rybicki}, G.~B. 1988, \apj, 335, 914,
  \dodoi{10.1086/166977}

\bibitem[{{Owocki} {et~al.}(1994){Owocki}, {Cranmer}, \& {Blondin}}]{Owocki94}
{Owocki}, S.~P., {Cranmer}, S.~R., \& {Blondin}, J.~M. 1994, \apj, 424, 887,
  \dodoi{10.1086/173938}

\bibitem[{{Proga}(1999)}]{Proga99b}
{Proga}, D. 1999, \mnras, 304, 938, \dodoi{10.1046/j.1365-8711.1999.02408.x}

\bibitem[{{Proga}(2007)}]{P07}
---. 2007, \apj, 661, 693, \dodoi{10.1086/515389}

\bibitem[{{Proga} \& {Kallman}(2004)}]{PK04}
{Proga}, D., \& {Kallman}, T.~R. 2004, \apj, 616, 688, \dodoi{10.1086/425117}

\bibitem[{{Proga} {et~al.}(1998){Proga}, {Stone}, \& {Drew}}]{Proga98}
{Proga}, D., {Stone}, J.~M., \& {Drew}, J.~E. 1998, \mnras, 295, 595,
  \dodoi{10.1046/j.1365-8711.1998.01337.x}

\bibitem[{{Proga} {et~al.}(1999){Proga}, {Stone}, \& {Drew}}]{Proga99a}
---. 1999, \mnras, 310, 476, \dodoi{10.1046/j.1365-8711.1999.02935.x}

\bibitem[{{Proga} {et~al.}(2000){Proga}, {Stone}, \& {Kallman}}]{PSK00}
{Proga}, D., {Stone}, J.~M., \& {Kallman}, T.~R. 2000, \apj, 543, 686,
  \dodoi{10.1086/317154}

\bibitem[{{Puls} {et~al.}(2000){Puls}, {Springmann}, \& {Lennon}}]{Puls00}
{Puls}, J., {Springmann}, U., \& {Lennon}, M. 2000, \aaps, 141, 23,
  \dodoi{10.1051/aas:2000312}

\bibitem[{{Quera-Bofarull} {et~al.}(2021){Quera-Bofarull}, {Done}, {Lacey},
  {Nomura}, \& {Ohsuga}}]{Quera-Bofarull21}
{Quera-Bofarull}, A., {Done}, C., {Lacey}, C.~G., {Nomura}, M., \& {Ohsuga}, K.
  2021, arXiv e-prints, arXiv:2111.02742.
\newblock \doarXiv{2111.02742}

\bibitem[{{Shu}(1992)}]{Shu92}
{Shu}, F.~H. 1992, {The physics of astrophysics. Volume II: Gas dynamics.}

\bibitem[{{Sim} {et~al.}(2010){Sim}, {Proga}, {Miller}, {Long}, \&
  {Turner}}]{Sim10}
{Sim}, S.~A., {Proga}, D., {Miller}, L., {Long}, K.~S., \& {Turner}, T.~J.
  2010, \mnras, 408, 1396, \dodoi{10.1111/j.1365-2966.2010.17215.x}

\bibitem[{{Srianand} {et~al.}(2002){Srianand}, {Petitjean}, {Ledoux}, \&
  {Hazard}}]{Srianand02}
{Srianand}, R., {Petitjean}, P., {Ledoux}, C., \& {Hazard}, C. 2002, \mnras,
  336, 753, \dodoi{10.1046/j.1365-8711.2002.05792.x}

\bibitem[{{Stevens} \& {Kallman}(1990)}]{SK90}
{Stevens}, I.~R., \& {Kallman}, T.~R. 1990, \apj, 365, 321,
  \dodoi{10.1086/169486}

\bibitem[{{Stone} \& {Proga}(2009)}]{Stone09}
{Stone}, J.~M., \& {Proga}, D. 2009, \apj, 694, 205,
  \dodoi{10.1088/0004-637X/694/1/205}

\bibitem[{{Stone} {et~al.}(2020){Stone}, {Tomida}, {White}, \&
  {Felker}}]{Stone20}
{Stone}, J.~M., {Tomida}, K., {White}, C.~J., \& {Felker}, K.~G. 2020, \apjs,
  249, 4, \dodoi{10.3847/1538-4365/ab929b}

\bibitem[{{Trigo} {et~al.}(2013){Trigo}, {Boirin}, {Migliari}, {Miller-Jones},
  {Parmar}, \& {Sidoli}}]{Trigo13b}
{Trigo}, M.~D., {Boirin}, L., {Migliari}, S., {et~al.} 2013, in Feeding Compact
  Objects: Accretion on All Scales, ed. C.~M. {Zhang}, T.~{Belloni},
  M.~{M{\'e}ndez}, \& S.~N. {Zhang}, Vol. 290, 25--28

\bibitem[{{Wang} {et~al.}(2022){Wang}, {Yang}, {Bu}, \& {Huang}}]{Wang22}
{Wang}, B.-C., {Yang}, X.-H., {Bu}, D.-F., \& {Huang}, S.-S. 2022, \mnras, 515,
  5594, \dodoi{10.1093/mnras/stac2203}

\bibitem[{{Waters} \& {Proga}(2016)}]{WP16}
{Waters}, T., \& {Proga}, D. 2016, \mnras, 460, L79,
  \dodoi{10.1093/mnrasl/slw056}

\bibitem[{{Waters} {et~al.}(2021){Waters}, {Proga}, \& {Dannen}}]{Waters21}
{Waters}, T., {Proga}, D., \& {Dannen}, R. 2021, \apj, 914, 62,
  \dodoi{10.3847/1538-4357/abfbe6}

\end{thebibliography}
\end{document}